\documentclass[aps,pra,preprint,tightenlines,showpacs,amsmath,amssymb]{revtex4-1}
\usepackage{graphicx}
\usepackage{bm}
\usepackage{slashed}
\usepackage{color}

\newcommand{\matheur}[1]{\text{\fontencoding{U}\fontfamily{eur}\fontseries{m}\fontshape{n}\selectfont#1}}
\newcommand{\matheurb}[1]{\text{\fontencoding{U}\fontfamily{eur}\fontseries{b}\fontshape{n}\selectfont#1}}
\newcommand{\op}[1]{\hat{\matheur{#1}}} 
\newcommand{\opb}[1]{\hat{\matheurb{#1}}} 
\newcommand{\dr}[1]{\mathrm{d}#1\,}
\newcommand{\ddr}[1]{\frac{\mathrm{d}}{\mathrm{d}#1}}
\newcommand{\ii}{\mathrm{i}}  
\newcommand{\ee}{\mathrm{e}}  
\newcommand{\bra}[1]{\langle{#1}\vert}
\newcommand{\ket}[1]{\vert{#1}\rangle}
\newcommand{\scal}[2]{\langle #1 | #2 \rangle} 

\begin{document}

\title{Generalized eikonal approximation for strong-field ionization}
\author{F. Cajiao V\'elez}
\author{K. Krajewska}
\email[E-mail address:\;]{Katarzyna.Krajewska@fuw.edu.pl}
\author{J. Z. Kami\'nski}
\affiliation{Institute of Theoretical Physics, Faculty of Physics, University of Warsaw, Pasteura 5,
02-093 Warsaw, Poland\\ \email[E-mail address:\;]{Katarzyna.Krajewska@fuw.edu.pl}}
\date{\today}

\begin{abstract}
We develop the eikonal perturbation theory to describe the strong-field ionization by finite laser pulses.
This approach in the first order with respect to the binding potential (the so-called generalized eikonal approximation) 
avoids a singularity at the potential center. Thus, in contrast to the ordinary eikonal approximation,
it allows to treat rescattering phenomena in terms of quantum trajectories. We demonstrate how
the first Born approximation and its domain of validity follow from eikonal perturbation
theory. Using this approach, we study the coherent diffraction patterns in photoelectron
energy spectra and their modifications induced by the interaction of photoelectrons with the
atomic potential. Along with these first results, we discuss the prospects of using the 
generalized eikonal approximation to study strong-field ionization from multi-centered atomic 
systems and to study other strong-field phenomena.

\end{abstract}


\maketitle

\section{Introduction}
\label{sec::intro}

Historically, the eikonal approximation was first introduced in relation to light scattering~\cite{Bruns,Born}.
Since then, such theory has been extensively applied not only in optics but also in atomic and molecular physics, quantum field theory, high energy physics, etc. 
The first application to modern theories was the analysis of scattering processes of energetic particles, as originally proposed by Moli\`{e}re~\cite{eikhist1}.
The idea behind the Moli\`{e}re approximation was the interpretation of particle trajectories as classical, straight line paths, which can be explained provided 
that the de Broglie wavelength of the scattered particle is small compared with the size of the scatterer~\cite{Joachain,bib:burke,Landau}. 
The pioneering work of Glauber~\cite{eikhist2} extended the Moli\`{e}re eikonal to the interaction of fast particles with complex atomic 
and nuclear systems, and established the basis of an exceptionally valuable technique to treat scattering processes at high energies.

After the theoretical basis were set, the eikonal approximation (EA) was successfully applied to more complex systems. For example, it was applied for the treatment 
of the inverse bremsstralung heating, which involves the electron scattering by a potential in the presence of intense electromagnetic fields~\cite{eiklas1,eiklas2}. 
In the original work of Choudhury and Bakar~\cite{eiklas1}, and of Zon~\cite{eiklas2}, the scattering amplitude was obtained by using the EA 
for the nonrelativistic Schr\"odinger equation and the laser field was considered as a monochromatic plane wave. The dipole approximation 
was extensively used in such calculations. Kristi\'{c} and Mittleman~\cite{bib:krstic} have shown that, for ionization in strong laser fields, 
the electron motion should be treated relativistically and the dipole approximation is not good enough to determine correctly the electron energy 
distribution and the transition rate as a function of the laser intensity. In order to account for those facts, the systematic eikonal 
perturbation theory was formulated by Kami\'nski~\cite{Jurek} for the nonrelativistic Schr\"odinger equation and the relativistic Klein-Gordon equation. 
For nonrelativistic case the dipole approximation for the laser pulse was used and applied to free-free transitions, whereas for the relativistic one 
the finite, in general, laser pulses in the plane-wave fronted approximation were considered. Note that the approach developed by Kami\'nski in~\cite{Jurek}
was based on the proper-time method, which for propagators in quantum theories was proposed by Fock~\cite{Fock}, and further developed 
by Schwinger~\cite{Schwinger,Schwinger2}, and by Fradkin and co-workers~\cite{Fradkin1,Fradkin2,Fradkin3} (see also the review \cite{bib:kleber}).

It has been demonstrated that, for potential scattering in the presence of strong fields, the EA
has certain limitations. Specifically, such a theory cannot be applied in regions far from the potential interaction zone~\cite{geiksch}. 
Other considerations restrict the EA just to moderate laser intensities~\cite{geikdirac}. The so-called generalized eikonal approximation (GEA), 
which is the first order term of the eikonal perturbation theory~\cite{Jurek}, overcame part of the mentioned problems by including certain quantum properties of the system in the eikonal limit.
The inclusion of such terms extended the range of applicability of the theory to large distances 
from the interaction site~\cite{geiksch} allowing a proper treatment of the scattering amplitude. The GEA was further extended to solving the 
relativistic Dirac equation in order to account approximately for the electron spin effects~\cite{geikdirac}.

In Refs.~\cite{Reis,Krainov}, the so-called eikonal-Volkov approximation was introduced to treat strong-field ionization of atoms. 
This consisted in including the laser field in full extent by means of the Volkov wave function, together with the ionic potential  
treated within the EA (see, for instance, also Refs.~\cite{Krai,Gord,Goresla,Faisal1,Faisal2,Potvliege,Thumm}). This approach was further developed
by Smirnova {\it et al.}~\cite{Smirnova1,Smirnova} to describe molecular ionization. 
It included the dynamical analysis of the electrons according to complex trajectories, which is a powerful 
tool for the overall understanding of the process (for complex trajectory method, see, for instance, recent reviews~\cite{Becker,Popruzh}).

Another way to treat the photoionization by intense laser fields is the so-called strong field approximation (SFA), sometimes recognized as the Keldysh-Faisal-Reiss 
or KFR theory~\cite{Keldysh,Faisal,Reiss}. The principal idea of the SFA approach is that, after the photoelectron appears in the continuum, it is treated as a free particle 
interacting with the laser field, ignoring any interaction with the nuclear potential during its excursion. Even though the SFA theory 
is known as one of the most fruitful analytical approaches in strong-field physics, its application can be justified for short range potentials only, 
not for the Coulomb type potentials present during ionization of neutral atoms or positively charged ions (see, also the most recent review by Popruzhenko~\cite{Popruzh}). 
The SFA theory gives, in general, a very good qualitative 
agreement with experimental and numerical results for laser light of moderate intensities in interaction with a short range potential~\cite{bib:wbecker}
but it fails to account for several well-documented spectral characteristics, including the wrong predictions obtained for the total 
ionization rate in the static field limit~\cite{Popruzhenko}.

Various methods have been proposed for taking the Coulomb interaction into account within the SFA. Except of the aforementioned eikonal-Volkov 
approach~\cite{Reis,Krainov,Krai,Gord,Goresla,Faisal1,Faisal2,Potvliege,Thumm,Smirnova1,Smirnova,Klaiber}, it has been done, for instance, by means of the Coulomb-Volkov anzatz~\cite{Tzoar,Cavaliere,CK0,CK1,CK2,CK3} 
which accounts for the asymptotic phase of the atomic field-free wave function. In this context also, the saddle point method was reformulated in terms 
of quantum trajectories by Popov and co-workers~\cite{Perelomov,Popov1,Popov2,Popov3}, 
by Gribakin and Kuchiev~\cite{Gribakin}, and reviewed recently by Popruzhenko~\cite{Popruzh}. The quantum trajectories
together with a detailed analysis of the saddle point equations, have shown to make considerable improvements 
in the SFA to include the Coulomb interactions~\cite{Popruzhen}. In our paper, we use the quantum trajectories as presented in Ref.~\cite{Popruzhenko}. 
Note that the extension of this method to the Dirac equation has been also reported~\cite{eikHeidelberg}.

In the present paper we propose a generalization of the eikonal approximation, in order to analyze the photoionization of atoms or ions 
by short laser pulses. This approach avoids a singularity at the center of the binding potential and, in contrast to the EA, 
can be analyzed in terms of quantum trajectories. In Sec.~\ref{sec::theory}, we set the physical and mathematical basis of the GEA
and derive an analytical expression for the ionization amplitude including the binding potential interaction. We show how 
the EA is obtained as a special case of the GEA for short times or for large distances from the potential center. In the same way, we demonstrate 
that the first Born approximation can be directly derived from our more general approach. Sec.~\ref{combs} 
is devoted to the numerical analysis of the photoelectron spectra generated by hydrogen interacting with an intense short pulse. 
Using both the Keldysh theory and the GEA theory, we discuss the coherent diffraction pattern in the photoelectron energy spectra and the role played by the atomic potential.
We present our concluding remarks in Sec.~\ref{conclusions}.

\section{Theoretical formulation}
\label{sec::theory}

For a time-dependent problem described by the Hamiltonian $\op{H}(t)$, which for our
further purpose we separate into two parts,
\begin{equation}
 \op{H}(t)=\op{H}_1(t)+\op{H}_2(t),
\label{t1}
\end{equation}
the time-evolution operator $\op{U}(t,t')$ satisfies the following Schr\"odinger equation,
\begin{equation}
\ii\ddr{t}\op{U}(t,t')=\op{H}(t)\op{U}(t,t'),
\label{t2}
\end{equation}
with the initial condition, $\op{U}(t',t')=\op{I}$. Here, $\op{I}$ is the identity operator.
The solution to the above equation, which incorporates also the initial condition, can be written in the form
\begin{equation}
\op{U}(t,t')=\op{T}\exp\Bigl(-\ii\int_{t'}^t \dr{\tau}\op{H}(\tau)\Bigr),
\label{t3}
\end{equation}
where $\op{T}$ is the time-ordering operator~\cite{Fetter,Landau}. Let us also introduce operators
$\op{U}_1(t,t')$ and $\op{U}_2(t,t')$ which determine the time-evolution governed by
the Hamiltonians $\op{H}_1(t)$ and $\op{H}_2(t)$, respectively. In other words, for each of them
it happens that
\begin{equation}
\ii\ddr{t}\op{U}_i(t,t')=\op{H}_i(t)\op{U}_i(t,t'), \quad \op{U}_i(t',t')=\op{I},
\label{t4}
\end{equation}
and
\begin{equation}
\op{U}_i(t,t')=\op{T}\exp\Bigl(-\ii\int_{t'}^t \dr{\tau}\op{H}_i(\tau)\Bigr), 
\label{t5}
\end{equation}
for $i=1,2$. It can be shown that the propagator $\op{U}(t,t')$ given by Eq.~\eqref{t3},
fulfills the integral Lippmann-Schwinger equation such that
\begin{equation}
\op{U}(t,t')=\op{U}_1(t,t')-\ii\int_{t'}^t \dr{\tau}\op{U}(t,\tau)\op{H}_2(\tau)\op{U}_1(\tau,t').
\label{t6}
\end{equation}
Since now on, we will assume that the Hamiltonian $\op{H}_1(t)$ is independent of time, 
$\op{H}(t)\equiv\op{H}_1$, whereas the Hamiltonian $\op{H}_2(t)$ varies with time in the
interval when $t\in[0,T]$ and is zero otherwise,
\begin{equation}
 \op{H}_2(t)=0 \quad \textrm{for}\quad t<0\quad \textrm{and}\quad t>T.
\label{t7}
\end{equation}
Our aim is to calculate the probability amplitude for a system governed by the
Hamiltonian $\op{H}(t)$ to make a transition from the initial state $\ket{\psi_{\mathrm{i}}}$ 
to the final state $\ket{\psi_{\mathrm{f}}}$. We assume that these are stationary states 
of the Hamiltonian $\op{H}_1$,
\begin{equation}
 \op{H}_1\ket{\psi_{\mathrm{f,i}}}=E_{\mathrm{f,i}}\ket{\psi_{\mathrm{f,i}}},
\label{t8}
\end{equation}
which are orthogonal $\langle\psi_{\mathrm{f}}|\psi_{\mathrm{i}}\rangle=0$. For completeness,
we note that their time-evolution is given by
\begin{equation}
 \ket{\psi_{\mathrm{f,i}}(t)}=\op{U}_1(t,0)\ket{\psi_{\mathrm{f,i}}}=\ee^{-\ii E_{\mathrm{f,i}}t}\ket{\psi_{\mathrm{f,i}}}.
\label{t9}
\end{equation}
The aforementioned probability amplitude calculated at time $T$ is
\begin{equation}
{\cal A}_{\mathrm{f,i}}(T)=\bra{\psi_{\mathrm{f}}(T)}\op{U}(T,0)\ket{\psi_{\mathrm{i}}(0)}.
\label{t10}
\end{equation}
It follows from Eq.~\eqref{t6} that this quantity can be rewritten as
\begin{equation}
{\cal A}_{\mathrm{f,i}}(T)=\scal{\psi_{\mathrm{f}}(T)}{\psi_{\mathrm{i}}(T)}
-\ii \int_0^T \dr{t'} \bra{\psi_{\mathrm{f}}(T)}\op{U}(T,t')\op{H}_2(t')\ket{\psi_{\mathrm{i}}(t')},
\label{t11}
\end{equation}
where the first term vanishes due to the orthogonality of the initial and final states.

\subsection{Ionization probability amplitude}
\label{amplitude}

The above theory can be conveniently applied to describe short-pulse ionization processes.
For this purpose, let us specify that $\op{H}_1$ is the atomic Hamiltonian,
\begin{equation}
 \op{H}_1=\frac{1}{2m}\opb{p}^2+V(\opb{r}),\label{t12}
\end{equation}
whereas $\op{H}_2$ describes the coupling with the laser field which, in the length gauge, has the form
\begin{equation}
 \op{H}_2=-e{\bm{\mathcal{E}}}(t)\cdot\opb{r}.\label{t13}
\end{equation}
At this point, we also specify that the vector potential, which describes a finite laser pulse,
${\bm A}(t)$ depends on time for $t\in[0,T]$ and is 0 otherwise. It is related to the electric 
field $\bm{\mathcal{E}}(t)$ such that ${\bm A}(t)=-\int_0^t {\rm d}\tau \bm{\mathcal{E}}(\tau)$.
The initial state is the atomic ground state of energy $E_0$, $\psi_0({\bm r})$,
which evolves in time according to
\begin{equation}
 \bra{{\bm r}}{\psi_{\rm i}(t)}\rangle=\ee^{-\ii E_0t}\psi_0({\bm r}).
\label{t14}
\end{equation}
The final state is the scattering state $\psi_{{\bm p}}^{(-)}({\bm r})$ which describes
a particle of momentum ${\bm p}$, 
\begin{equation}
 \bra{{\bm r}}{\psi_{\rm f}(t)}\rangle=\exp\Bigl(-\ii \frac{{\bm p}^2}{2m}t\Bigr)\psi_{{\bm p}}^{(-)}({\bm r}).
\label{t15}
\end{equation}
Next, let us define the retarded propagator in the length gauge $K_L({\bm r},t; {\bm r}',t')$ which relates to the quantum-mechanical dynamics governed by the
Hamiltonian $\op{H}(t)$,
\begin{equation}
 K_L({\bm r},t; {\bm r}',t')=\theta(t-t')\bra{{\bm r}}\op{U}(t,t')\ket{{\bm r}'},
\label{t17}
\end{equation}
with the initial condition that
\begin{equation}
 K_L({\bm r},t+0; {\bm r}',t)=\delta({\bm r}-{\bm r}').
\label{t18}
\end{equation}
With these definitions, the transition probability amplitude~\eqref{t11} for ionization becomes
\begin{equation}
{\cal A}({\bm p})=-\ii \int_0^T\dr t'\exp\Bigl(\ii\frac{{\bm p}^2}{2m}T-\ii E_0t'\Bigr)\int{\rm d}^3r\int{\rm d}^3r'
\psi_{{\bm p}}^{(-)*}({\bm r})K_L({\bm r},T;{\bm r}',t')(-e\bm{\mathcal{E}}(t')\cdot {\bm r}')\psi_0({\bm r}').
\label{t19}
\end{equation}
Therefore, for our further analysis it is necessary to derive an explicit form of the propagator 
$K_L({\bm r},t;{\bm r}',t')$ for an electron under a simultaneous action of the laser field and the external potential. 
This will be done in the next Section, using eikonal perturbation theory~\cite{Jurek}.

\subsection{Retarded propagator under the GEA}
\label{propagator}

While our ultimate goal is to derive the retarded propagator in the length gauge, we start with the velocity gauge
and the respective propagator $K_V({\bm r},t;{\bm r}',t')$. For the time being we consider the most general case
in which the scalar potential also depends on time, $V({\bm r},t)$. In the present case, $K_V({\bm r},t;{\bm r}',t')$ satisfies Eq.~\eqref{t17}
but with the Hamiltonian written in the velocity gauge,
\begin{equation}
\Bigl(-\ii\frac{\partial}{\partial t'}-\frac{1}{2m}[\ii{\bm \nabla}'-e{\bm A}(t')]^2-V({\bm r}',t')
\Bigr)K_V({\bm r},t;{\bm r'},t')=\ii\delta(t-t')\delta({\bm r}-{\bm r'}).
\label{t20}
\end{equation}
Then, one can look for the propagator $K_V({\bm r},t;{\bm r}',t')$ using the Fock-Schwinger {\it proper-time} representation~\cite{Jurek},
\begin{align}
K_V({\bm r},t;{\bm r'},t')=\int_0^\infty {\rm d}s \int\frac{{\rm d}\Omega\,{\rm d}^3k}{(2\pi)^4}\exp\Big[-&\ii\Omega(t-t')
+\ii{\bm k}\cdot({\bm r}-{\bm r}')+\ii s\bigl(\Omega-\frac{{\bm k}^2}{2m}+\ii\varepsilon\bigr) \nonumber \\
+&\ii\Phi_{{\bm k}}(t',s)+\ii\chi_{\bm k}({\bm r}',t',s)\Big],  
\label{t21}
\end{align}
with $s$ being the proper time. Here, we have introduced unknown functions $\Phi_{\bm k}(t',s)$ and $\chi_{\bm k}({\bm r}',t',s)$.
These functions must satisfy the following conditions
\begin{eqnarray}
\Phi_{\bm k}(t',s)&=&0\quad {\rm when}\quad  {\bm A}={\bm 0},\label{t22}\\
\chi_{\bm k}(\bm{r}',t',s)&=&0\quad {\rm when}\quad V=0,\label{t23}
\end{eqnarray}
in which case the propagator~\eqref{t21} becomes a free particle propagator~\cite{Jurek,bib:kleber}. When
substituting the formula~\eqref{t21} into Eq.~\eqref{t20} we arrive at
\begin{align}
\label{t23a}
\int \frac{\mathrm{d}\Omega\mathrm{d}^3k}{(2\pi)^4}&\ee^{-\ii\Omega(t-t')+\ii \bm{k}\cdot (\bm{r}-\bm{r}')}\int_0^{\infty}\mathrm{d}s \Bigl[\Omega-\frac{\bm{k}^2}{2m}
+\frac{e}{m}\bm{k}\cdot\bm{A}(t')-\frac{e^2}{2m}\bm{A}^2(t')-V(\bm{r}',t') \\
+&\partial_{t'}\Phi_{\bm{k}}+\partial_{t'}\chi_{\bm{k}}+\frac{1}{m}(\bm{k}-e\bm{A}(t'))\cdot\bm{\nabla}'\chi_{\bm{k}}+\frac{\ii}{2m}\Delta'\chi_{\bm{k}}-\frac{1}{2m}\bigl(\bm{\nabla}'\chi_{\bm{k}}\bigr)^2\Bigr] \nonumber \\
\times & \exp\Bigl[\ii s\Bigl(\Omega-\frac{\bm{k}^2}{2m}+\ii\varepsilon\Bigr)+\ii\Phi_{\bm{k}}+\ii\chi_{\bm{k}}\Bigr]=\ii\delta(t-t')\delta(\bm{r}-\bm{r}'). \nonumber
\end{align}
If $\Phi_{\bm k}(t',0)=0$ and $\chi_{\bm k}({\bm r}',t',0)=0$, the equation above is fulfilled provided that
\begin{equation}
\label{t23b}
\int_0^{\infty}\mathrm{d}s\,\partial_s\exp\Bigl[\ii s\Bigl(\Omega-\frac{\bm{k}^2}{2m}+\ii\varepsilon\Bigr)+\ii\Phi_{\bm{k}}+\ii\chi_{\bm{k}}\Bigr]=\ii\, .
\end{equation}
As a result, we obtain a partial differential
equation for functions $\Phi_{\bm k}(t',s)$ and $\chi_{\bm k}({\bm r}',t',s)$, which then can be separated
into two independent equations:
\begin{align}
\Bigl(\frac{\partial}{\partial t'}-\frac{\partial}{\partial s}\Bigr)\Phi_{\bm k}(t',s)=&-\frac{e}{m}{\bm A}(t')\cdot\bigl[{\bm k}-\frac{e}{2}{\bm A}(t')\bigr],
\label{t24}\\
\label{t25}
\Bigl(\frac{\partial}{\partial t'}-\frac{\partial}{\partial s}\Bigr)\chi_{\bm k}({\bm r}',t',s)=&
-\frac{1}{m}\bigl[{\bm k}-e{\bm A}(t')\bigr]\cdot{\bm \nabla}'\chi_{\bm k}({\bm r}',t',s)+V({\bm r}',t') \\
+&\frac{1}{2m}\bigl({\bm \nabla}'\chi_{\bm k}({\bm r}',t',s)\bigr)^2-\frac{\ii}{2m}\Delta'\chi_{\bm k}({\bm r}',t',s) \nonumber . 
\end{align}
When solving these equations one has to remember of the above initial conditions for $s=0$ and about Eqs.~\eqref{t22} and~\eqref{t23}.
It is rather straightforward to solve Eq.~\eqref{t24}. On the other hand, Eq.~\eqref{t25} is a nonlinear second order differential equation
for $\chi_{\bm k}({\bm r}',t',s)$, which can be solved explicitly only for particular potentials, for instance, for the harmonic oscillator. 
Since now on we will call $\chi_{\bm k}({\bm r}',t',s)$ the {\it eikonal}.

It follows from Eq.~\eqref{t24} that
\begin{equation}
\Phi_{\bm k}(t',s)=\int_{t'}^{t'+s}{\rm d}\tau\; \frac{e}{m}{\bm A}(\tau)\cdot\bigl[{\bm k}-\frac{e}{2}{\bm A}(\tau)\bigr].
\label{t26}
\end{equation}
Taking this into account we conclude that in the absence of external potential (in which case $\chi_{\bm k}=0$), the exact propagator~\eqref{t21}
reduces to the Volkov propagator. In order to solve Eq.~\eqref{t25} perturbatively, let us first rewrite this equation such that
\begin{equation}
 \Bigl(\frac{\partial}{\partial t'}-\frac{\partial}{\partial s}+\frac{1}{m}\bigl[{\bm k}-e{\bm A}(t')\bigr]\cdot {\bm \nabla}'
+\frac{\ii}{2m}\Delta'\Bigr)\chi_{\bm k}({\bm r}',t',s)=W_{\bm k}({\bm r}',t',s),
\label{t27}
\end{equation}
where we have defined,
\begin{equation}
W_{\bm k}({\bm r}',t',s)= V({\bm r}',t')+\frac{1}{2m}\bigl({\bm \nabla}'\chi_{\bm k}({\bm r}',t',s)\bigr)^2.
\label{t28}
\end{equation}
Note that in the case when $V({\bm r}',t')=0$, it follows from Eqs.~\eqref{t23} and~\eqref{t28} that $W_{\bm k}({\bm r}',t',s)=0$.
Now, with the help of the Fourier transforms,
\begin{eqnarray}
 \chi_{\bm k}({\bm r}',t',s)&=&\int\frac{{\rm d}^3\mu}{(2\pi)^3}\;\ee^{\ii{\bm \mu}\cdot{\bm r}'}\tilde\chi_{\bm k}({\bm \mu},t',s),
\label{t29}\\
 W_{\bm k}({\bm r}',t',s)&=&\int\frac{{\rm d}^3\mu}{(2\pi)^3}\;\ee^{i{\bm \mu}\cdot{\bm r}'}\tilde W_{\bm k}({\bm \mu},t',s),
\label{t30}
\end{eqnarray}
we replace Eq.~\eqref{t27} by 
\begin{equation}
\Bigl(\frac{\partial}{\partial t'}-\frac{\partial}{\partial s}+\frac{\ii{\bm \mu}}{m}\cdot[{\bm k}-e{\bm A}(t')]-\frac{\ii{\bm \mu}^2}{2m}\Bigr)\tilde\chi_{\bm k}({\bm \mu},t',s)=\tilde W_{\bm k}({\bm \kappa},t',s).
\label{t31}
\end{equation}
We look for the solution of this equation in the form,
\begin{equation}
\tilde\chi_{\bm k}({\bm \mu},t',s)=\exp\Bigl[\ii\int_{t'}^{t'+s}{\rm d}\tau\Bigl(\frac{{\bm \mu}}{m}\cdot[{\bm k}-e{\bm A}(\tau)]-\frac{{\bm \mu}^2}{2m}\Bigr)\Bigr]\tilde\chi_{\bm k}'({\bm \mu},t',s).
\label{t32}
\end{equation}
By putting this solution into Eq.~\eqref{t31}, we find out that a new function $\tilde\chi_{\bm k}'({\bm \mu},t',s)$
satisfies the following equation,
\begin{equation}
\Bigl(\frac{\partial}{\partial t'}-\frac{\partial}{\partial s}\Bigr)\tilde\chi_{\bm k}'({\bm \mu},t',s)=\tilde W_{\bm k}({\bm \mu},t',s)\exp\Bigl[-\ii\int_{t'}^{t'+s}{\rm d}\tau\Bigl(\frac{{\bm \mu}}{m}\cdot[{\bm k}-e{\bm A}(\tau)]-\frac{{\bm \mu}^2}{2m}\Bigr)\Bigr],
\label{t33}
\end{equation}
which leads to
\begin{equation}
 \tilde\chi_{\bm k}'({\bm \mu},t',s)=-\int_{t'}^{t'+s}{\rm d}\sigma\,\tilde W_{\bm k}({\bm \mu},\sigma,t'+s-\sigma)
\exp\Bigl[-\ii\int_{\sigma}^{t'+s}{\rm d}\tau \Bigl(\frac{{\bm \mu}}{m}\cdot[{\bm k}-e{\bm A}(\tau)]-\frac{{\bm \mu}^2}{2m}\Bigr)\Bigr].
\label{t34}
\end{equation}
From here, it also follows that
\begin{align}
\tilde\chi_{\bm k}({\bm \mu},t',s)=-\exp\Bigl[\ii\int_{t'}^{t'+s}{\rm d}\tau & \Bigl(\frac{{\bm \mu}}{m}\cdot[{\bm k}-e{\bm A}(\tau)]-\frac{{\bm \mu}^2}{2m}\Bigr)\Bigr] 
\int_{t'}^{t'+s}{\rm d}\sigma\,\tilde W_{\bm k}({\bm \mu},\sigma,t'+s-\sigma) \nonumber \\
\times & \exp\Bigl[-\ii\int_\sigma^{t'+s}{\rm d}\sigma'\Bigl(\frac{{\bm \mu}}{m}\cdot[{\bm k}-e{\bm A}(\sigma')]
-\frac{{\bm \mu}^2}{2m}\Bigr)\Bigr]. \label{t35}
\end{align}

Before we proceed further, we introduce new quantities
\begin{align}
{\bm a}_{\bm k}(t)=&\frac{1}{m}\int_0^t {\rm d}\tau [{\bm k}-e{\bm A}(\tau)],\label{t36}\\
{\bm R}_{\bm k}({\bm r}',t',\sigma)=&{\bm r}'+{\bm a}_{\bm k}(\sigma)-{\bm a}_{\bm k}(t')
={\bm r}'+\frac{1}{m}\int_{t'}^\sigma {\rm d}\tau [{\bm k}-e{\bm A}(\tau)]. \label{t45}
\end{align}
Then, Eq.~\eqref{t35} can be rewritten in a more compact form,
\begin{equation}
\tilde\chi_{\bm k}({\bm \mu},t',s)=-\int_{t'}^{t'+s}{\rm d}\sigma\,\tilde W_{\bm k}({\bm \mu},\sigma,t'+s-\sigma)
\exp\Bigl(\ii{{\bm \mu}}\cdot[{\bm a}_{\bm k}(\sigma)-{\bm a}_{\bm k}(t')]-\ii\frac{{\bm \mu}^2}{2m}(\sigma-t')\Bigr), \label{t37}
\end{equation}
which, after substituting into Eq.~\eqref{t29}, gives
\begin{align}
\chi_{\bm k}({\bm r}',t',s)=&-\int_{t'}^{t'+s}{\rm d}\sigma\int {\rm d}^3 \rho\,W_{\bm k}({\bm \rho},\sigma,t'+s-\sigma) \nonumber \\
\times & \int\frac{{\rm d}^3\mu}{(2\pi)^3}\exp\Bigl(\ii{\bm \mu}\cdot[{\bm R}_{\bm k}({\bm r}',t',\sigma)-{\bm \rho}]-\frac{\ii{\bm\mu}^2}{2m}(\sigma-t')\Bigr).
\label{t38}
\end{align}
Here, we recognize that the integral over ${\bm \mu}$ is the Fresnel integral,
and so it can be performed exactly. In doing so, we arrive at the following expression for the eikonal,
\begin{align}
 \chi_{\bm k}({\bm r}',t',s)=&-\int_{t'}^{t'+s}{\rm d}\sigma\int {\rm d}^3 \rho\,\Bigl(\frac{m}{2\pi\ii (\sigma-t')}\Bigr)^{3/2}\exp\Bigl(\frac{\ii m}{2(\sigma-t')}[{\bm R}_{\bm k}({\bm r}',t',\sigma)-{\bm \rho}]^2\Bigr)\nonumber\\
\times & W_{\bm k}({\bm \rho},\sigma,t'+s-\sigma)\equiv -\int_{t'}^{t'+s}{\rm d}\sigma\,V_{\rm eff}\bigl({\bm R}_{\bm k}({\bm r}',t',\sigma),t',\sigma\bigr),
\label{t39}
\end{align}
which implicitly defines an effective potential $V_{\rm eff}$. This is the starting point for the eikonal perturbation theory and for the GEA.

For completeness, let us go back to Eq.~\eqref{t21} and perform the respective integrals over 
$\Omega$ and $s$. As a result, we obtain the integral representation of the retarded
propagator in the velocity gauge such that
\begin{align}
K_V({\bm r},t;{\bm r'},t')=\int\frac{{\rm d}^3k}{(2\pi)^3}\exp\Big[&\ii{\bm k}\cdot({\bm r}-{\bm r}')
-\ii (t-t')\frac{{\bm k}^2}{2m}\label{t40}\\
+&\ii\Phi_{{\bm k}}(t',t-t')+\ii\chi_{\bm k}({\bm r}',t',t-t')\Big],  \nonumber
\end{align}
with the functions $\Phi_{{\bm k}}(t',s)$ and $\chi_{\bm k}({\bm r}',t',s)$ defined by Eqs.~\eqref{t26} and~\eqref{t39},
respectively. Since
\begin{equation}
\frac{{\bm k}^2}{2m}(t-t')-\Phi_{{\bm k}}(t',t-t')=\frac{m}{2}\int_{t'}^t {\rm d}\sigma\Bigl(\frac{\partial {\bm R}_{\bm k}({\bm r}',t',\sigma)}{\partial\sigma}\Bigr)^2
\label{p1}
\end{equation}
we can rewrite Eq.~\eqref{t40} such that
\begin{align}
K_V({\bm r},t;{\bm r'},t')=& \int\frac{{\rm d}^3k}{(2\pi)^3}\exp\Bigl\{\ii{\bm k}\cdot({\bm r}-{\bm r}')\label{p2}\\
-&\ii \int_{t'}^t {\rm d}\sigma\Bigl[\frac{m}{2}\Bigl(\frac{\partial {\bm R}_{\bm k}({\bm r}',t',\sigma)}{\partial\sigma}\Bigr)^2
+V_{\rm eff}\bigl({\bm R}_{\bm k}({\bm r}',t',\sigma),t',\sigma\bigr)\Bigr]\Bigr\}.  \nonumber
\end{align}
On the other hand, one can show that in the length gauge the retarded propagator has the following form,
\begin{align}
K_L({\bm r},t;{\bm r'},t')=\int\frac{{\rm d}^3k}{(2\pi)^3}\exp\Big[&\ii({\bm k}-e{\bm A}(t))\cdot{\bm r}
-\ii({\bm k}-e{\bm A}(t'))\cdot {\bm r}'-\ii (t-t')\frac{{\bm k}^2}{2m}\nonumber\\
+&\ii\Phi_{{\bm k}}(t',t-t')+\ii\chi_{\bm k}({\bm r}',t',t-t')\Big],  
\label{t41}
\end{align}
with exactly same functions $\Phi_{{\bm k}}(t',s)$ and $\chi_{\bm k}({\bm r}',t',s)$ as before. This, in turn, leads to
\begin{align}
K_L({\bm r},t;{\bm r'},t')=\int\frac{{\rm d}^3k}{(2\pi)^3}\exp\Bigl\{&\ii m\frac{\partial {\bm R}_{\bm k}({\bm r}',t',t)}{\partial t}\cdot{\bm r}
+\ii m\frac{\partial {\bm R}_{\bm k}({\bm r}',t',t)}{\partial t'}\cdot{\bm r}' \label{p3} \\
-&\ii \int_{t'}^t {\rm d}\sigma\Bigl[\frac{m}{2}\Bigl(\frac{\partial {\bm R}_{\bm k}({\bm r}',t',\sigma)}{\partial\sigma}\Bigr)^2+V_{\rm eff}\bigl({\bm R}_{\bm k}({\bm r}',t',\sigma),t',\sigma\bigr)\Bigr]\Bigr\}, \nonumber
\end{align}
and is the most general form of the propagator defining the ionization amplitude in Eq.~\eqref{t18}.

In the first order approximation with respect to the potential, $W_{\bm k}({\bm r}',t',s)$ can be approximated by $V({\bm r}',t')$
which is the essence of the GEA. Hence, we can write that, in the first approximation, the eikonal~\eqref{t39} becomes
\begin{equation}
 \chi_{\bm k}^{(1)}({\bm r}',t',s)=-\int_{t'}^{t'+s}{\rm d}\sigma\,V_{\rm eff}^{(1)}\bigl({\bm R}_{\bm k}({\bm r}',t',\sigma),t',\sigma\bigr),
\label{p4}
\end{equation}
where
\begin{align}
 V_{\rm eff}^{(1)}\bigl({\bm R}_{\bm k}({\bm r}',t',\sigma),t',\sigma\bigr)=&\int {\rm d}^3 \rho\, \Bigl(\frac{m}{2\pi\ii (\sigma-t')}\Bigr)^{3/2} \label{t42}  \\
\times &\exp\Bigl(\frac{\ii m}{2(\sigma-t')}[ {\bm R}_{\bm k}({\bm r}',t',\sigma)-{\bm \rho}]^2\Bigr)V({\bm \rho},\sigma).
\nonumber
\end{align}
We will also use the respective notation for the propagators $K_V^{(1)}({\bm r},t;{\bm r'},t')$ and $K_L^{(1)}({\bm r},t;{\bm r'},t')$. 
Note that for the static potential, when $V({\bm \rho},\sigma)=V({\bm \rho})$, $V_{\mathrm{eff}}^{(1)}$ depends explicitly only on $\bm{R}_{\bm k}$ and $\sigma -t'$. 

At this point, let us investigate the limit of short time intervals, i.e.,
when $\sigma\approx t'$; in this context the short-time limit is equivalent with the classical one, when $\hbar\rightarrow 0$. To this end, in Eq.~\eqref{t42} we make use of the following model of the delta function
\begin{equation}
\delta({\bm r})=\lim_{\varepsilon\rightarrow 0}\frac{\exp\Bigl(\displaystyle \ii\frac{{\bm r}^2}{2\varepsilon}\Bigr)}{(2\pi\ii\varepsilon)^{3/2}},
\label{delta}
\end{equation}
which leads to $\delta\bigl({\bm R}_{\bm k}({\bm r}',t',\sigma)-{\bm \rho}\bigr)$. Performing the remaining
spatial integral in~\eqref{t42}, we arrive at the conclusion that 
\begin{equation}
 V_{\rm eff}^{(1)}\bigl({\bm R}_{\bm k}({\bm r}',t',\sigma),t',\sigma\bigr)\approx V({\bm R}_{\bm k}({\bm r}',t',\sigma),\sigma),
\label{limit}
\end{equation}
which holds in the limit when $\sigma\approx t'$. This result is in full agreement with the Dirac conjecture \cite{Dirac,Dirac2} that in 
the short-time limit (and only in this limit, i.e., when the quantum spreading of the electron wave packet is negligible) 
the propagator is proportional to $\ee^{\ii S_{\mathrm{cl}}}$, where $S_{\mathrm{cl}}$ is the classical action.
This afterward has lead Feynman to the path integrals~\cite{FeynmanHibbs}.

In closing this Section, let us also note that the integral equation for the eikonal~\eqref{t39}
allows to construct a series expansion with respect to the potential -- the eikonal
perturbation theory. Note that such theory and the GEA, as its first order term, was proposed in Ref.~\cite{Jurek} for both the nonrelativistic Schr\"odinger 
and relativistic Klein-Gordon equations, and it was applied to free-free transitions. In this paper we
further extend it for nonrelativistic ionization processes.

\subsection{Generalized eikonal for the Coulomb potential}
\label{coulomb}

For the Coulomb potential describing the interaction of an electron of charge 
$e<0$ and a nucleus of charge $-Ze$, where $Z=1,2,...$ is the atomic number, $V({\bm r},t)\equiv V({\bm r})=-Z\alpha c/r$.
Here, $\alpha=e^2/(4\pi\varepsilon_0 c)$ is the fine structure constant; in the atomic units used in our numerical analysis $\alpha c=1$. The eikonal defined by Eq.~\eqref{p4} becomes
\begin{align}
 \chi_{\bm k}^{(1)}({\bm r}',t',s)=&\int_{t'}^{t'+s}{\rm d}\sigma\int {\rm d}^3 \rho\Bigl(\frac{m}{2\pi\ii(\sigma-t')}\Bigr)^{3/2}\nonumber\\
\times&\frac{Z\alpha c}{\rho}\exp\Bigl(\frac{\ii m}{2(\sigma-t')}[ {\bm R}_{\bm k}({\bm r}',t',\sigma)-{\bm \rho}]^2\Bigr),
\label{t43}
\end{align}
where the integral over ${\bm \rho}$ can be performed exactly. This leads to 
\begin{equation}
\chi_{\bm k}^{(1)}({\bm r'},t',s)=Z\alpha c\int_{t'}^{t'+s}{\rm d}\sigma\,\frac{1}{\left|{\bm R}_{\bm k}({\bm r}',t',\sigma)\right|}{\rm erf}\Bigl(\sqrt{\frac{m}{2\ii (\sigma-t')}}\,|{\bm R}_{\bm k}({\bm r}',t',\sigma)|\Bigr),    
\label{t44}
\end{equation}
where ${\rm erf}(z)$ is the error function.
The commonly used eikonal (see, for instance, in~\cite{eikhist1,eikhist2,Jurek,Smirnova} and references therein) which, contrary to the above approximation,
is singular for the Coulomb potential,
\begin{equation}
\chi_{{\bm k},{\rm original}}({\bm r}',t',s)=Z\alpha c\int_{t'}^{t'+s}{\rm d}\sigma\,\frac{1}{\left|{\bm R}_{\bm k}({\bm r}',t',\sigma)\right|},
\label{t46}
\end{equation}
is recovered, if
\begin{equation}
 \sqrt{\frac{m}{2(\sigma-t')}}\,|{\bm R}_{\bm k}({\bm r}',t',\sigma)|\gg 1.
\label{t47}
\end{equation}
Note that this condition is in agreement with the short-time interval approximation, $\sigma\approx t'$, which has been discussed above. 
However, for $\sigma\not\approx t'$, the aforementioned condition requires that $\left|{\bm R}_{\bm k}({\bm r}',t',\sigma)\right| \not\rightarrow 0$.
In other words, the approach proposed in this paper is also applicable to cases when the electron trajectory can return 
back to the origin of the Coulomb potential. For this reason, it does not lead to problems already mentioned in Ref.~\cite{Popruzhenko}. 
In our case: (i) the integral in \eqref{t44} for $\sigma$ close to $t'$ converges (for the integrand we meet only 
the integrable $(\sigma-t')^{-1/2}$ singularity), and (ii) the generalized eikonal is not singular for the zeroth-order trajectories 
that may revisit the nucleus in real time. It is worth mentioning that the condition~\eqref{t47} is fulfilled for large 
distances from the Coulomb center. This means that the original and generalized eikonals coincide with each other not only 
for short times, but also at distant points in space. Hence, if the electron wave packets or quantum trajectories are 
far away from the center during the time evolution both approximations should give similar results. This is usually the case 
if the final kinetic energy of photoelectrons is much larger than $3U_p$, where $U_p$ is the ponderomotive energy defined below 
[Eq.~\eqref{a8}]. This will be demonstrated later on.

In closing this Section we note that similar close expressions for the eikonal $\chi_{\bm k}^{(1)}({\bm r}',t',s)$ can be also derived for other potentials such as
the Yukawa, Gaussian or multi-center Coulomb potentials. These cases will be studied in due course.

\subsection{Ionization probability amplitude in the GEA}
\label{eikonal}

In the first order eikonal approximation, the ionization probability amplitude in the length gauge equals
\begin{align}
{\cal A}^{(1)}({\bm p})=&-\ii \int_0^T\dr t'\exp(\ii\frac{{\bm p}^2}{2m}T-\ii E_0t')\nonumber\\
\times & \int{\rm d}^3r\int{\rm d}^3r'
\psi_{{\bm p}}^{(-)*}({\bm r})K_L^{(1)}({\bm r},T;{\bm r}',t')(-e\bm{\mathcal{E}}(t')\cdot {\bm r}')\psi_0({\bm r}'),
\label{t48}
\end{align}
with
\begin{align}
K_L^{(1)}({\bm r},T;{\bm r'},t')=\int\frac{{\rm d}^3k}{(2\pi)^3}\exp\Big[ & \ii{\bm k}\cdot{\bm r}-\ii({\bm k}-e{\bm A}(t'))\cdot {\bm r}'
-\ii (T-t')\frac{{\bm k}^2}{2m}\nonumber\\
+&\ii\Phi_{{\bm k}}(t',T-t')
+\ii\chi_{\bm k}^{(1)}({\bm r}',t',T-t')\Big]. 
\label{t49}
\end{align}
Having this in mind and performing the integral over ${\bm r}$ in Eq.~\eqref{t48}, we arrive at
\begin{align}
\label{t50}
{\cal A}^{(1)}&({\bm p})=-\ii \int_0^T\dr t'\exp(\ii\frac{{\bm p}^2}{2m}T-\ii E_0t')\int{\rm d}^3r'\int\frac{{\rm d}^3k}{(2\pi)^3}\,\tilde\psi_{{\bm p}}^{(-)*}({\bm k})(-e\bm{\mathcal{E}}(t')\cdot {\bm r}')\psi_0({\bm r}')\\
\times&\exp\Bigl[-\ii({\bm k}-e{\bm A}(t'))\cdot {\bm r}'-\ii (T-t')\frac{{\bm k}^2}{2m}+\ii\Phi_{{\bm k}}(t',T-t')+\ii\chi_{\bm k}^{(1)}({\bm r}',t',T-t')\Bigr]. \nonumber
\end{align}
In the following, we assume that the Fourier transform of the final scattering state, $\tilde\psi_{{\bm p}}^{(-)}({\bm k})$,
is centered around ${\bm k}\sim{\bm p}$. Essentially, this corresponds to a plane wave approximation in the final electron state which is
the common approximation applied in strong-field physics.
In other words $|\psi_{\bm p}^{(-)*}({\bm 0})|$, which emerges under this approximation in Eq.~\eqref{t50}, contributes only to the normalization of the plane wave. Therefore, we can disregard this
multiplication factor remembering that the density of final electron states per unit volume equals ${\rm d}^3p/(2\pi)^3$. 
This way, the probability amplitude of ionization~\eqref{t50} becomes
\begin{align}
 {\cal A}^{(1)}({\bm p})=&-\ii\int_0^T\dr t'\exp\Bigl[\ii\bigl(\frac{{\bm p}^2}{2m}-E_0\bigr)t'+\ii\Phi_{{\bm p}}(t',T-t')\Bigr] \label{t51} \\
\times & \int{\rm d}^3r' (-e\bm{\mathcal{E}}(t')\cdot {\bm r}')\psi_0({\bm r}')
\exp\Bigl[-\ii({\bm p}-e{\bm A}(t'))\cdot {\bm r}'+\ii\chi_{\bm p}^{(1)}({\bm r}',t',T-t')\Bigr], \nonumber
\end{align}
where $\Phi_{{\bm p}}(t',T-t')$ and $\chi_{\bm p}^{(1)}({\bm r}',t',T-t')$ are defined by Eqs.~\eqref{t26} and~\eqref{t44}, respectively.

\subsection{Limit of the Born approximation}
\label{bornapp}

In order to establish relations between the GEA and the Born approximation, we go back to Eq.~\eqref{t19}. As in the previous Section, we assume 
that the final scattering state is described by the plane wave of momentum $\bm{p}$, i.e., $\psi^{(-)}_{\bm{p}}(\bm{r})\approx \ee^{\ii\bm{p}\cdot\bm{r}}$. 
The retarded Volkov propagator in the length gauge, $K_L^{(0)}(\bm{r},t;\bm{r}',t')$, can be obtained from Eq.~\eqref{t41} by neglecting $\chi_{\bm{k}}$. 
Hence,
\begin{align}
\label{born1}
K_L^{(0)}(\bm{r},t;\bm{r}',t')=&\int\frac{\mathrm{d}^3k}{(2\pi)^3}\exp\Bigl[\ii(\bm{k}-e\bm{A}(t))\cdot\bm{r} \\
&-\ii(\bm{k}-e\bm{A}(t'))\cdot\bm{r}'-\ii\int_{t'}^t\mathrm{d}\tau\frac{1}{2m}\bigl(\bm{k}-e\bm{A}(\tau)\bigr)^2
\Bigr]. \nonumber
\end{align}
The wave function $\psi_{\bm{p}}^{(0)}(\bm{r}',t')$, defined by the integral
\begin{align}
\label{born2}
\psi_{\bm{p}}^{(0)*}(\bm{r}',t')=&\int\mathrm{d}^3r \ee^{-\ii\bm{p}\cdot\bm{r}}K_L^{(0)}(\bm{r},T;\bm{r}',t') \\
=&\exp\Bigl[-\ii(\bm{p}-e\bm{A}(t'))\cdot\bm{r}'-\ii\int_{t'}^T\mathrm{d}\tau\frac{1}{2m}\bigl(\bm{p}-e\bm{A}(\tau)\bigr)^2 \Bigr], \nonumber
\end{align}
is the Volkov solution of the Schr\"odinger equation in the length gauge. One can check that it fulfills the boundary condition
\begin{equation}
\label{born3}
 \psi_{\bm{p}}^{(0)}(\bm{r}',T)=\ee^{\ii\bm{p}\cdot\bm{r}'},
\end{equation}
as for $t'\geqslant T$ the action of the laser pulse vanishes.

The exact retarded propagator, $K_L(\bm{r},t;\bm{r}',t')$, satisfies the Lippmann-Schwinger equation in accordance with Eq.~\eqref{t6},
\begin{align}
\label{born4}
K_L(\bm{r},t;\bm{r}',t')=&K_L^{(0)}(\bm{r},t;\bm{r}',t') \\
-&\ii\int_{t'}^t\mathrm{d}\tau \int\mathrm{d}^3y K_L^{(0)}(\bm{r},t;\bm{y},\tau)V(\bm{y},\tau)K_L(\bm{y},\tau;\bm{r}',t'), \nonumber
\end{align}
which allows to split the exact probability amplitude [cf. Eq.~\eqref{t19}] into two terms,
\begin{equation}
\mathcal{A}(\bm{p})=\mathcal{A}_{K}(\bm{p})+\mathcal{A}_{\mathrm{resc}}(\bm{p}),
\label{born5}
\end{equation}
where
\begin{equation}
\label{born6}
\mathcal{A}_{K}(\bm{p})=-\ii\int_0^T\mathrm{d}t'\exp\Bigl[\ii\frac{\bm{p}^2}{2m}T-\ii E_0t'\Bigr] 
\int\mathrm{d}^3r' \psi_{\bm{p}}^{(0)*}(\bm{r}',t')(-e\bm{\mathcal{E}}(t')\cdot\bm{r}')\psi_0(\bm{r}') 
\end{equation}
is the Keldysh amplitude, and
\begin{align}
\label{born7}
\mathcal{A}_{\mathrm{resc}}(\bm{p})=-\int_0^T\mathrm{d}t'\exp\Bigl[\ii\frac{\bm{p}^2}{2m}T-\ii E_0t'\Bigr] & \int\mathrm{d}^3r'\int_{t'}^T\mathrm{d}\tau\int\mathrm{d}^3y \psi_{\bm{p}}^{(0)*}(\bm{y},\tau)V(\bm{y},\tau)  \\
\times & K_L(\bm{y},\tau;\bm{r}',t')(-e\bm{\mathcal{E}}(t')\cdot\bm{r}')\psi_0(\bm{r}') \nonumber
\end{align}
is the exact rescattering amplitude. In the first Born approximation [by the Born approximation we understand the expansion with respect to the potential $V({\bm r},t)$]
the exact propagator in the above equation is replaced by the Volkov propagator in order to get the amplitude that describes the rescattering of electrons after ionization. 
After some algebraic manipulations one can show that
\begin{align}
\label{born8}
K_L^{(0)}(\bm{y},\tau;\bm{r}',t')=&\frac{\psi_{\bm{p}}^{(0)*}(\bm{r}',t')}{\psi_{\bm{p}}^{(0)*}(\bm{y},\tau)} \Bigl(\frac{m}{2\pi\ii (\tau-t')}\Bigr)^{3/2} \\
&\times\exp\Bigl[\frac{\ii m}{2 (\tau-t')}\Bigl(\bm{R}_{\bm{p}}(\bm{r}',t',\tau)-\bm{y}\Bigr)^2\Bigr]. \nonumber
\end{align}
Inserting this identity into \eqref{born7}, we can write down that the rescattering amplitude in the first Born approximation is 
$\mathcal{A}_{\mathrm{resc}}(\bm{p})\approx \mathcal{A}_{\mathrm{B1}}(\bm{p})$, where
\begin{align}
\label{born9}
\mathcal{A}_{\mathrm{B1}}(\bm{p})=&\int\mathrm{d}^3r'\int_0^T\mathrm{d}t'\exp\Bigl[\ii\frac{\bm{p}^2}{2m}T-\ii E_0t'\Bigr] \\
&\times\psi_{\bm{p}}^{(0)*}(\bm{r}',t')\chi_{\bm{p}}^{(1)}(\bm{r}',t',T-t')(-e\bm{\mathcal{E}}(t')\cdot\bm{r}')\psi_0(\bm{r}') \nonumber
\end{align}
and $\chi_{\bm{p}}^{(1)}(\bm{r}',t',s)$ is defined by Eqs.~\eqref{p4} and \eqref{t42}. It follows from Eq.~\eqref{t51} that 
the total probability amplitude in the first Born approximation, $\mathcal{A}_{K}(\bm{p})+\mathcal{A}_{\mathrm{B1}}(\bm{p})$, is exactly recovered from 
our first order GEA provided that $|\chi_{\bm{p}}^{(1)}(\bm{r}',t',T-t')|\ll 1$. This means that one can use the Taylor expansion
\begin{equation}
\label{born10}
\ee^{\ii\chi_{\bm{p}}^{(1)}(\bm{r}',t',T-t')}\approx 1+\ii\chi_{\bm{p}}^{(1)}(\bm{r}',t',T-t').
\end{equation}
Such an agreement is not achievable within the EA. This shows that the first Born approximation, which has been extensively used 
in the analysis of rescattering processes in ionization~\cite{Becker}, is the limiting case of the generalized eikonal 
expansion.

\section{Combs in the photoelectron spectrum}
\label{combs}

\subsection{Laser pulse and its characteristics}
\label{laser}

In the above formulation we have assumed that a finite laser pulse lasts for time $T$ and, therefore, it is described by the electric field $\bm{\mathcal{E}}(t)$
which vanishes for $t<0$ and $t>T$. The pulse duration $T$ defines the fundamental frequency of field oscillations, $\omega=2\pi/T$. One can also introduce
the field phase, $\phi=\omega t$, which allows to rewrite the above condition such that $\bm{\mathcal{E}}(\phi)$ vanishes for $\phi<0$ and $\phi>2\pi$.
We assume that the driving pulse is linearly polarized along the $z$-axis. In the dipole approximation, the laser field is described by the electric field vector
\begin{equation}
\bm{\mathcal{E}}(\phi)={\cal E}_0 f_{\cal E}(\phi){\bm e}_z,
\label{a2}
\end{equation}
where ${\cal E}_0$ is related to the amplitude of field oscillations. Here, the shape function $f_{\cal E}(\phi)$ is adjusted such that $f_{{\cal E}}(\phi)=0$ for $\phi<0$ and $\phi>2\pi$, and it has to satisfy the condition~\cite{Becker}
\begin{equation}
\int_0^{2\pi}f_{\cal E}(\phi){\rm d}\phi=0.
\label{a3}
\end{equation}
This condition is fulfilled provided that the shape function has the following Fourier decomposition,
\begin{equation}
f_{\cal E}(\phi)=\sideset{}{'}\sum_{N=-N_0}^{N_0}{\cal E}_N\ee^{-\ii N\phi},
\label{a4}
\end{equation}
where $\sum'$ means that the zeroth Fourier component is excluded from the sum, $N\neq 0$. Since $f_{\cal E}(\phi)$ is a real function,
we also require that ${\cal E}_N^*={\cal E}_{-N}$. This expansion allows us to define the average intensity carried out by the laser pulse,
\begin{equation}
I=\bigl\langle c\varepsilon_0 {\mathcal{\bm{E}}}^2\bigr\rangle=\frac{1}{2\pi}\int_0^{2\pi}c\varepsilon_0{\mathcal{\bm{E}}}^2(\phi){\rm d}\phi.
\label{inten1}
\end{equation}
Namely,
\begin{equation}
I=2c\varepsilon_0{\cal E}_0^2\sum_{N=1}^{N_0}|{\cal E}_N|^2.
\label{inten2}
\end{equation}
Note that this definition, even though derived for a finite laser pulse, is consistent with the monochromatic plane wave approximation.
In the latter case, taking the electric field of the form $\bm{\mathcal{E}}(\phi)={\cal E}_0 \sin(\phi){\bm e}_z$,
we have ${\cal E}_N=\mp\ii/2$ for $N=\pm 1$. Hence, it follows from Eq.~\eqref{inten2} that the averaged intensity of the monochromatic plane wave equals
$I=c\varepsilon_0{{\cal E}_0}^2/2$.

Let us now define the shape function for the vector potential ${\bm A}(\phi)$,
\begin{equation}
f_A(\phi)=-\int_0^\phi f_{\cal E}(\varphi){\rm d}\varphi,
\label{a5}
\end{equation}
which leads to
\begin{equation}
{\bm A}(\phi)=\frac{{\cal E}_0}{\omega}f_A(\phi){\bm e}_z.
\label{a6}
\end{equation}
Using Eq.~\eqref{a4} one can derive the corresponding Fourier decomposition of the shape function $f_A(\phi)$ [Eq.~\eqref{a5}]
\begin{equation}
f_A(\phi)=A_0+\sideset{}{'}\sum_{N=-N_0}^{N_0}A_N\ee^{-\ii N\phi},
\label{a7}
\end{equation}
where $\displaystyle A_0=-2\sum_{N=1}^{N_0}{\rm Re}A_N$ and $A_N=-\ii {\cal E}_N/N$ assuming that $N\neq 0$. Since we have imposed the condition ${\bm A}(0)={\bm A}(2\pi)={\bm 0}$, the vector potential
has a constant and an oscillatory contributions, ${\bm A}(\phi)={\bm A}_{\rm const}+{\bm A}_{\rm osc}(\phi)$.
It is the oscillatory contribution to the vector potential, ${\bm A}_{\rm osc}(\phi)$, which describes the quiver motion of free electrons in the laser field.
The ponderomotive energy of such a motion can be defined as
\begin{equation}
U_p=\bigl\langle\frac{e^2{\bm A}_{\rm osc}^2}{2m}\bigr\rangle=\frac{1}{2\pi}\int_0^{2\pi}\frac{e^2}{2m}{\bm A}_{\rm osc}^2(\phi){\rm d}\phi.
\label{a8}
\end{equation}
Applying here the series expansion~\eqref{a7}, we find out that 
\begin{equation}
U_p=\frac{e^2{\cal E}_0^2}{m\omega^2}\sum_{N=1}^{N_0}\frac{|{\cal E}_N|^2}{N^2}.
\label{a9}
\end{equation}
Again, for the monochromatic plane wave, we obtain from Eq.~\eqref{a9} that $U_p=e^2{\cal E}_0^2/(4m\omega^2)$. This is the well-known 
formula for the ponderomotive energy of a free electron driven by the monochromatic plane wave. 

For our further purpose, we introduce also the vector function
\begin{equation}
\bm{\alpha}(\phi)=\frac{\mathcal{E}_0}{\omega^2}f_{\alpha}(\phi){\bm e}_z,
\label{a10p}
\end{equation}
where
\begin{equation}
f_{\alpha}(\phi)=-\int_0^{\phi}f_A(\varphi)\mathrm{d}\varphi=f_{\alpha,0}+f_{\alpha,1}\phi+f_{\alpha,\mathrm{osc}}(\phi),
\label{a10}
\end{equation}
with
\begin{align}
f_{\alpha,0}&=2\sum_{N=1}^{N_0}\frac{\mathrm{Re}\mathcal{E}_N}{N^2},
\label{a11a} \\
f_{\alpha,1}&=2\sum_{N=1}^{N_0}\frac{\mathrm{Im}\mathcal{E}_N}{N},
\label{a11b} \\
f_{\alpha,\mathrm{osc}}(\phi)&=-\sideset{}{'}\sum_{N=-N_0}^{N_0}\frac{\mathrm{Re}\mathcal{E}_N}{N^2}\ee^{-\ii N\phi}.
\label{a11c} 
\end{align}
This function will be used in Sec.~\ref{trajectories}.

In this paper, we consider the laser field described by the shape function
\begin{equation}
f_{\cal E}(\phi)=\begin{cases}\sin^2(N_{\rm rep}\frac{\phi}{2})\sin(N_{\rm rep}\phi), & \phi\in [0,2\pi], \cr 0, & {\rm otherwise}. \end{cases}
\label{a12}
\end{equation}
Such a laser field consists of $N_{\rm rep}$ single-cycle pulses with no time delay in-between (where $N_{\rm rep}=1,2,3,...$). 
Introducing the laser frequency $\omega_{\rm L}=N_{\rm rep}\,\omega$, we can represent Eq.~\eqref{a12} as
\begin{equation}
f_{\cal E}(t)=\begin{cases}\frac{1}{2}\sin(\omega_{\rm L}t)-\frac{1}{4}\sin(2\omega_{\rm L}t), & t\in [0,T], \cr 0, & {\rm otherwise}. \end{cases}
\label{a13}
\end{equation}
This clearly shows that the laser field~\eqref{a12} can be composed out of two harmonics, i.e., $\omega_{\rm L}$ and $2\omega_{\rm L}$. For our choice of
the shape function, the only nonzero coefficients in its Fourier expansion~\eqref{a4} are ${\cal E}_{\pm N_{\rm rep}}=\mp\ii/4$ and ${\cal E}_{\pm 2N_{\rm rep}}=\pm\ii/8$.
Therefore, according to Eq.~\eqref{inten2}, the averaged intensity carried out by the laser pulse~\eqref{a12} is
\begin{equation}
I=\frac{5}{32}c\varepsilon_0{\cal E}_0^2.
\end{equation}
Moreover, it follows from Eq.~\eqref{a9} that the ponderomotive energy associated with the quiver motion of an electron in such a field equals
\begin{equation}
U_p=\frac{17}{256}\,\frac{e^2{\cal E}_0^2}{m\omega_{\rm L}^2}.
\label{pond}
\end{equation}
Further we shall assume that $\omega_{\rm L}$
equals the frequency of the Ti-Sapphire laser, $\omega_{\rm L}=1.55$eV, while the averaged intensity of the pulse is $I=3.125\times 10^{13}$ W/cm$^2$. 
Thus, the ponderomotive energy of the electron oscillating in the laser pulse~\eqref{a12} equals $U_p=1.024\omega_{\rm L}$.

Below we shall analyze the energy spectra of photoelectrons ionized from a hydrogen-like atom by the laser field~\eqref{a12} with different $N_{\rm rep}$. 
This means that while changing the number of pulse repetitions, $N_{\rm rep}$,
the time duration of the entire sequence of pulses, $T=2\pi N_{\rm rep}/\omega_{\rm L}$, will change as well. First, we will present the respective results based on the Keldysh theory.

\subsection{Combs in the Keldysh theory}
\label{combsKeldysh}

In the Keldysh approximation, the amplitude of ionization ${\cal A}_K({\bm p})$ is given by Eq.~\eqref{born6}. This formula can be written explicitly in the form,
\begin{align}
&{\cal A}_K({\bm p})=-\ii\exp\Bigl\{\ii \Bigl(\frac{{\bm p}^2}{2m}-E_0\Bigr)T-\ii\int_0^T{\rm d}\tau\Bigl[\frac{1}{2m}\bigl({\bm p}-e{\bm A}(\tau)\bigr)^2-E_0\Bigr]\Bigr\}\int_0^T{\rm d}t'\int{\rm d}^3r'\nonumber\\
&\times\exp\Bigl\{-\ii ({\bm p}-e{\bm A}(t'))\cdot{\bm r}'+\ii\int_0^{t'}{\rm d}\tau\Bigl[\frac{1}{2m}\bigl({\bm p}-e{\bm A}(\tau)\bigr)^2-E_0\Bigr]\Bigr\}(-e\bm{\mathcal{E}}(t')\cdot\bm{r}')\psi_0(\bm{r}').
\label{a14}
\end{align}
Introducing here the phase of the laser pulse, $\phi$, and defining the following quantities:
\begin{eqnarray}
{\bm q}(\phi)&=&{\bm p}-e{\bm A}(\phi),\label{a15}\\
G({\bm p},\phi)&=&\frac{1}{\omega}\int_0^\phi {\rm d}\phi'\Bigl[\frac{{\bm q}^2(\phi')}{2m}-E_0\Bigr],\label{a16}
\end{eqnarray}
we can rewrite ${\cal A}_K({\bm p})$ as
\begin{eqnarray}
{\cal A}_K({\bm p})&=&-\ii\frac{\ee^{\ii\Phi_0({\bm p})}}{\omega}\int_0^{2\pi}{\rm d}\phi\,\ee^{\ii G({\bm p},\phi)}\int{\rm d}^3r'\ee^{-\ii{\bm q}(\phi)\cdot{\bm r}'}(-e\bm{\mathcal{E}}(\phi)\cdot\bm{r}')\psi_0(\bm{r}').
\label{a17}
\end{eqnarray}
Here, we have also introduced the following abbreviation, $\Phi_0({\bm p})=\bigl(\frac{{\bm p}^2}{2m}-E_0\bigr)T-G({\bm p},2\pi)$.

For a hydrogen-like atom in the ground state we have
\begin{equation}
\psi_0({\bm r})=\lambda\sqrt{\frac{\lambda}{\pi}}\ee^{-\lambda r},\quad E_0=-\frac{\lambda^2}{2m},
\label{a18}
\end{equation}
where $\lambda=(Za_0)^{-1}$ and $a_0$ is the Bohr radius. In this case the space integral in Eq.~\eqref{a17} can be performed exactly. As a result, we obtain
\begin{equation}
\int{\rm d}^3r'\ee^{-\ii{\bm q}(\phi)\cdot{\bm r}'}(-e\bm{\mathcal{E}}(\phi)\cdot\bm{r}')\psi_0(\bm{r}')=4\ii\lambda^2\frac{\sqrt{\lambda\pi}}{m^2\omega}\,
\frac{G''({\bm p},\phi)}{[G'({\bm p},\phi)]^3},
\label{a19}
\end{equation}
where, according to Eq.~\eqref{a16},
\begin{eqnarray}
G'({\bm p},\phi)&\equiv&\frac{\partial}{\partial\phi}G({\bm p},\phi)=\frac{1}{\omega}\Bigl[\frac{{\bm q}^2(\phi)}{2m}-E_0\Bigr],\label{a20}\\
G''({\bm p},\phi)&\equiv&\frac{\partial^2}{\partial\phi^2}G({\bm p},\phi)=\frac{1}{m\omega^2}\,e\bm{\mathcal{E}}(\phi)\cdot{\bm q}(\phi).\label{a21}
\end{eqnarray}
Inserting Eq.~\eqref{a19} into Eq.~\eqref{a17}, we obtain that the probability
amplitude of ionization of a hydrogen-like atom in the ground state, within the framework of the Keldysh theory, equals
\begin{equation}
{\cal A}_K({\bm p})=4\lambda^2\frac{\sqrt{\lambda\pi}}{m^2\omega^2}\,\ee^{\ii\Phi_0({\bm p})}
\int_0^{2\pi}{\rm d}\phi\frac{G''({\bm p},\phi)}{[G'({\bm p},\phi)]^3}\,\ee^{\ii G({\bm p},\phi)}.
\label{a22}
\end{equation}
To calculate the total probability of ionization, one has to integrate $|{\cal A}_K(\bm p)|^2$ 
over the density of final electron states, ${\rm d}^3p/(2\pi)^3$. In doing so, we arrive at the following formula,
\begin{equation}
{\cal P}_K=\int\frac{{\rm d}^3p}{(2\pi)^3}|{\cal A}_K({\bm p})|^2=\frac{m}{(2\pi)^3}\int{\rm d}\Omega_{\bm p}\int{\rm d}E_{\bm p}|{\bm p}| |{\cal A}_K({\bm p})|^2,
\label{a23}
\end{equation}
where $E_{\bm p}={\bm p}^2/(2m)$ is the electron kinetic energy and ${\rm d}\Omega_{\bm p}$ is the electron differential solid angle. 
The integral remaining in Eq.~\eqref{a22} can be performed numerically. Before proceeding with numerical calculations, let us derive also an approximate
formula for the probability amplitude~\eqref{a22} using the saddle point method. This analytic approach will help us later on to interpret our numerical results.

\subsubsection{Singular saddle point approximation}
\label{sec::saddle}

In light of Eq.~\eqref{a22} we consider the integral,
\begin{equation}
{\cal I}_\nu=\int_0^{2\pi}{\rm d}\phi\frac{H(\phi)}{[G'({\bm p},\phi)]^\nu}\,\ee^{\ii \xi G({\bm p},\phi)},
\label{a24}
\end{equation}
where $H(\phi)$ is a regular function of $\phi$ and $\xi>0$ is a large parameter. The saddle points $\phi_s$ are the solutions of the equation,
\begin{equation}
G'({\bm p},\phi)\bigl|_{\phi=\phi_s}=0.
\label{a25}
\end{equation}
While evaluating the integral in Eq.~\eqref{a22} at $\phi=\phi_s$, we encounter the problem in the denominator which contains $G'({\bm p},\phi)$.
Therefore, to avoid singularities at the saddle points, in Eq.~\eqref{a24} we substitute,
\begin{equation}
\frac{1}{[G'({\bm p},\phi)]^\nu}=\frac{1}{\Gamma(\nu)}\int_0^\infty{\rm d}\eta\,\eta^{\nu-1}\ee^{-G'({\bm p},\phi)\eta}.
\label{a26}
\end{equation}
Then, an ordinary saddle point method can be successfully applied which consists in replacing the function $H(\phi)$ by its value 
at the saddle point $\phi_s$ and, also, in replacing the argument of the exponent by the first nonvanishing terms arising from the Taylor expansion around $\phi_s$ \cite{Gribakin,Popruzh}.
After performing the remaining integrals we obtain that
\begin{equation}
{\cal I}_\nu\approx\frac{\pi\xi^{(\nu-1)/2}}{2^{(\nu-1)/2}\Gamma(\frac{\nu+1}{2})}\sum_s\ee^{\ii(\nu+1)\pi/4}\frac{H(\phi_s)\ee^{\ii \xi G({\bm p},\phi_s)}}{[G''({\bm p},\phi_s)]^{(\nu+1)/2}},
\label{a27}
\end{equation}
where the sum is over such saddle points that satisfy the conditions:
\begin{equation}
{\rm Im}[G({\bm p},\phi_s)]>0,\quad {\rm Im}[G''({\bm p},\phi_s)]>0.
\label{a28}
\end{equation}
As it will follow shortly, these conditions are compatible with the requirement that ${\rm Im}(\phi_s)>0$.

Going back to Eq.~\eqref{a22} and making use of Eq.~\eqref{a27} for $\nu=3$ we obtain that, under the saddle point approximation, the probability amplitude of photoionization
from the ground state of a hydrogen-like atom equals
\begin{equation}
{\cal A}_K^{(\rm saddle)}({\bm p})=-2\sqrt{\frac{\lambda}{\pi}}\Bigl(\frac{\pi\lambda}{m\omega}\Bigr)^2\,\ee^{\ii\Phi_0({\bm p})}\sum_s\frac{\ee^{\ii G({\bm p},\phi_s)}}{G''({\bm p},\phi_s)}.
\label{a29}
\end{equation}
This defines the total probability of ionization, in accordance with Eq.~\eqref{a23}. Because the contributing saddle points have to 
satisfy the conditions~\eqref{a28}, their careful analysis is necessary.

\begin{figure}
\includegraphics[width=7cm]{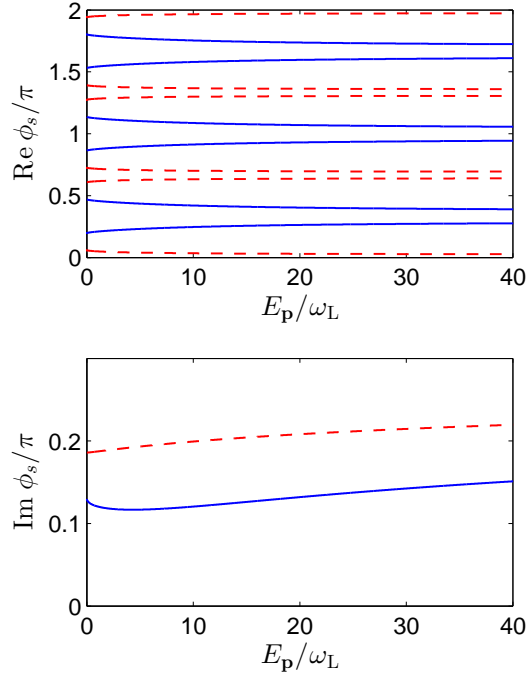}%
\caption{(Color online) Positions of the saddle points $\phi_s$ as a function of the kinetic energy of electrons $E_{\bm p}$, calculated from Eq.~\eqref{a25}.
Only these saddle points which satisfy~\eqref{a28} are plotted.
The saddle points with the same imaginary part are marked either as solid or dashed lines. The parameters of the driving laser field [described by
Eqs.~\eqref{a2} and~\eqref{a12}] are $\omega_{\rm L}=1.55$ eV, $N_{\rm rep}=3$, and $I=3.125\times 10^{13}$ W/cm$^2$. The final electrons are detected asymptotically 
at the polar angle $\theta_{\bm p}=0.2\pi$.
\label{PhaseSaddle20150106}}
\end{figure}

For our choice of the pulse shape~\eqref{a12}, there are in general $8N_{\rm rep}$ solutions of the equation~\eqref{a25}. However, a half of them does not fulfill the conditions~\eqref{a28}.
Among the remaining solutions we can distinguish two groups of solutions ($2N_{\rm rep}$ points each) with the exact same positive imaginary parts for $\phi_s$, $G({\bm p},\phi_s)$, 
and $G''({\bm p},\phi_s)$. To illustrate this, we consider a Ti:Sapphire laser ($\omega_{\rm L}=1.55$ eV) producing a field composed out
of three single-cycle pulses ($N_{\rm rep}=3$), with the electric field described by Eqs.~\eqref{a2} and~\eqref{a12}. We choose the averaged intensity in the pulse
$I=3.125\times 10^{13}$ W/cm$^2$. Due to cylindrical symmetry of our problem, the positions of saddle points do not depend on the azimuthal angle of ionized photoelectrons, 
just on their polar angle $\theta_{\bm p}$. Here, we choose $\theta_{\bm p}=0.2\pi$. In Fig.~\ref{PhaseSaddle20150106}, we plot the real (upper panel)
and imaginary (lower panel) parts of the solutions to Eq.~\eqref{a25} which obey the conditions~\eqref{a28}. For $N_{\rm rep}=3$, we observe 12 such saddle points.
The saddle points which are represented by the dashed line have larger imaginary parts than the saddle points which are represented by the solid line.
For this reason, the contribution of the former points to the sum in~\eqref{a27} is marginally small and can be disregarded in our further analysis.

\begin{figure}
\includegraphics[width=7cm]{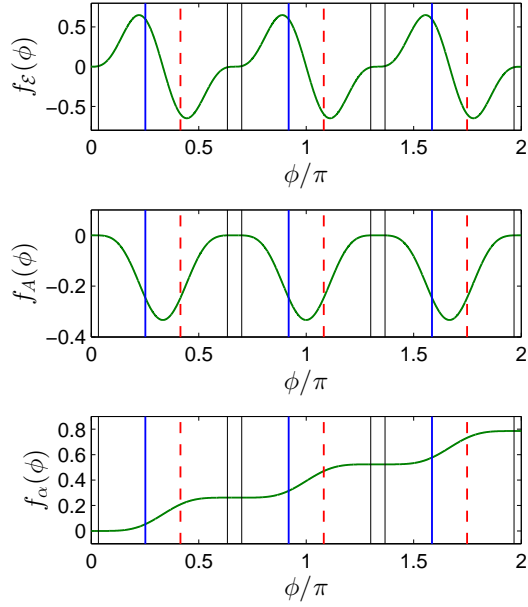}
\caption{(Color online) Shows the shape function $f_{\cal E}(\phi)$ for a triple ($N_{\rm rep}=3$) laser pulse defined by Eq.~\eqref{a12}. In the lower panels, 
the corresponding shape functions $f_A(\phi)$ and $f_\alpha(\phi)$ are displayed [Eqs.~\eqref{a5} and~\eqref{a10}, respectively]. We have marked the real
parts of saddle points, ${\rm Re}\phi_s$, as vertical lines. While the thin black lines correspond to the position of these saddle points which contribute 
very little to the probability amplitude of ionization~\eqref{a29}, the major contribution there comes from the saddle points marked as the thick 
(both solid and dashed) lines. The positions of ${\rm Re}\phi_s$ are for $E_{\bm p}\approx 12U_p$ and $\theta_{\bm p}=0.2\pi$.
\label{shapefunctionsandsaddlesv2}}
\end{figure}

In Fig.~\ref{shapefunctionsandsaddlesv2}, we present the shape functions $f_{\cal E}(\phi)$, $f_A(\phi)$, 
and $f_\alpha(\phi)$ for $N_{\rm rep}=3$. The vertical lines represent the real parts of saddle points, ${\rm Re}\phi_s$, for the electron kinetic 
energy $E_{\bm p}=12.22\omega_{\rm L}\approx 12U_p$.
The thin black lines correspond to those saddle points that do not contribute much to the probability amplitude of ionization~\eqref{a29}.
This is not surprising as their real parts correspond to the nearly zero value of the electric field.
On contrary, the remaining vertical lines (thick solid and dashed lines) correspond to the saddle points that have to be accounted for in Eq.~\eqref{a29}. 
We see that the important saddle points have their real parts which correspond to the nearly extreme values of the electric field.

\begin{figure}
\includegraphics[width=7cm]{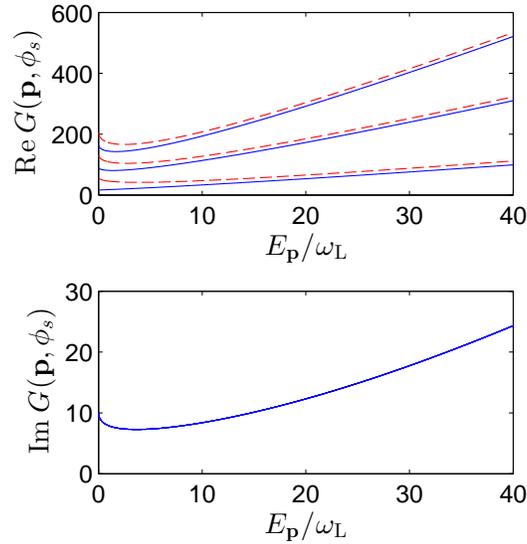}%
\caption{(Color online) Shows the dependence of the real (upper panel) and imaginary (lower panel) parts of $G({\bm p},\phi_s)$ [Eq.~\eqref{a16}]
on the photoionized electron kinetic energy, $E_{\bm p}$, calculated for the same parameters as in Fig.~\ref{PhaseSaddle20150106}. Only these
saddle points are accounted for which significantly contribute to the probability amplitude of ionization~\eqref{a29}.
\label{GSaddle20150106}}
\end{figure}
\begin{figure}
\includegraphics[width=7cm]{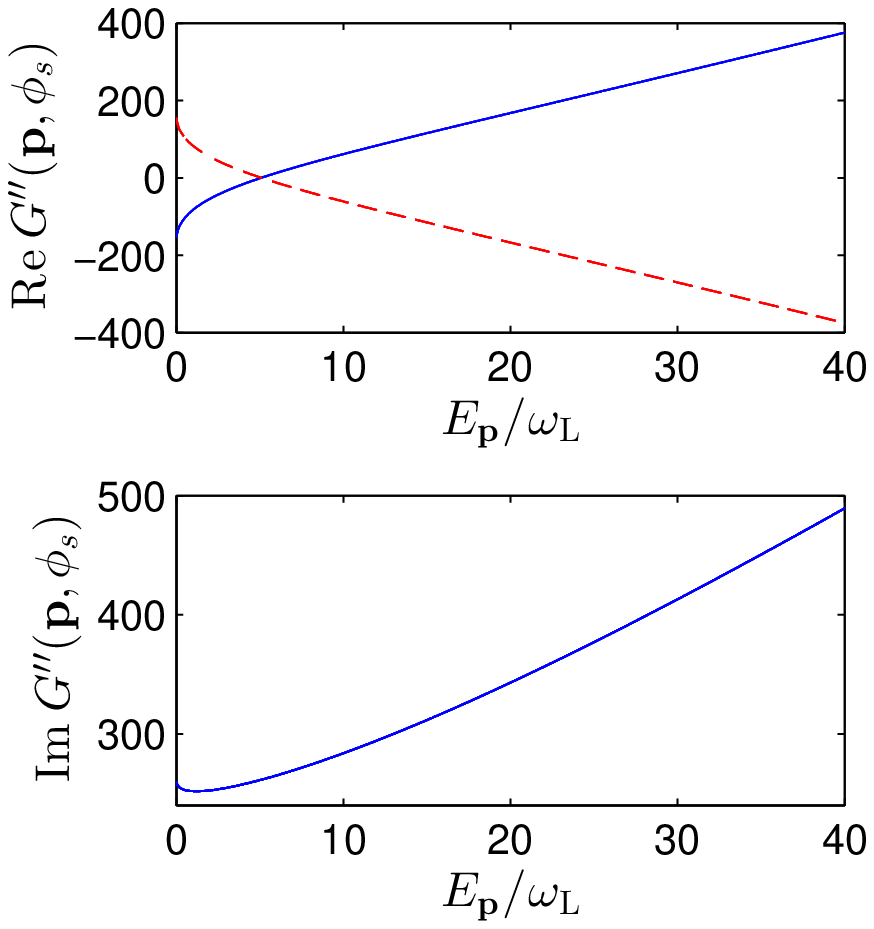}%
\caption{(Color online) The same as in Fig.~\ref{GSaddle20150106} but for the function $G''({\bm p},\phi_s)$.
\label{GBisSaddle20150106}}
\end{figure}

In Figs.~\ref{GSaddle20150106} and~\ref{GBisSaddle20150106}, we draw the dependence of the functions $G({\bm p},\phi_s)$ and
$G''({\bm p},\phi_s)$ on the kinetic energy of photoelectrons $E_{\bm p}$ for those saddle points $\phi_s$ that contribute
significantly to the probability amplitude of ionization~\eqref{a29}. These saddle points were denoted in Fig.~\ref{PhaseSaddle20150106}
by the solid lines. Among these points, we can distinguish between the ones that relate to the maxima (solid blue lines) and minima 
(dashed red lines) of the shape function $f_{\cal E}(\phi)$ (see, Fig.~\ref{shapefunctionsandsaddlesv2}).
We denote these saddle points as $\phi_{N_{\rm rep}}^{(\ell)}$ and $\tilde{\phi}_{N_{\rm rep}}^{(\ell)}$, respectively, with $\ell=1,2,...,N_{\rm rep}$.
They have the same positive imaginary part but their real parts differ such that
\begin{eqnarray}
{\rm Re}\phi_{N_{\rm rep}}^{(\ell)}&=&\phi_0+\frac{2\pi}{N_{\rm rep}}(\ell-1),\label{a30}\\
{\rm Re}\tilde{\phi}_{N_{\rm rep}}^{(\ell)}&=&-\phi_0+\frac{2\pi}{N_{\rm rep}}\ell.\label{a31}
\end{eqnarray}
Here, $\phi_0$ denotes the real part of the first saddle point which gives a significant contribution to the probability amplitude. 
It can be anticipated from the upper panel of Fig.~\ref{GSaddle20150106} that
\begin{eqnarray}
{\rm Re}[G({\bm p},\phi_{N_{\rm rep}}^{(\ell)})]&=&G_0({\bm p})+2\pi(\ell-1)F({\bm p}),\label{a32}\\
{\rm Re}[G({\bm p},\tilde{\phi}_{N_{\rm rep}}^{(\ell)})]&=&\tilde{G}_0({\bm p})+2\pi(\ell-1)F({\bm p}).\label{a33}
\end{eqnarray}
Even though it is possible to derive the exact forms of functions $G_0({\bm p}), \tilde{G}_0({\bm p})$, and $F({\bm p})$, it is not
of a particular interest. As we will show shortly, only the structure of the functions ${\rm Re}[G({\bm p},\phi_{N_{\rm rep}}^{(\ell)})]$ and ${\rm Re}[G({\bm p},\tilde{\phi}_{N_{\rm rep}}^{(\ell)})]$ 
is important for interpreting the resulting energy distributions of photoelectrons. Moreover, it follows from the bottom panel of Fig.~\ref{GSaddle20150106} that 
${\rm Im}[G({\bm p},\phi_{N_{\rm rep}}^{(\ell)})]={\rm Im}[G({\bm p},\tilde{\phi}_{N_{\rm rep}}^{(\ell)})]\equiv W({\bm p})>0$. Another important observation, based on Fig.~\ref{GBisSaddle20150106},
is that
\begin{eqnarray}
{\rm Re}[G''({\bm p},\phi_{N_{\rm rep}}^{(\ell)})]&=&-{\rm Re}[G''({\bm p},\tilde{\phi}_{N_{\rm rep}}^{(\ell)})],\label{a34}\\
{\rm Im}[G''({\bm p},\phi_{N_{\rm rep}}^{(\ell)})]&=&{\rm Im}[G''({\bm p},\tilde{\phi}_{N_{\rm rep}}^{(\ell)})].\label{a35}
\end{eqnarray}
Having this in mind, we shall denote in the following: $G_0''({\bm p})=|G''({\bm p},\phi_{N_{\rm rep}}^{(\ell)})|=|G''({\bm p},\tilde{\phi}_{N_{\rm rep}}^{(\ell)})|$ and
$\psi_{G''}({\bm p})={\rm arg}[G''({\bm p},\phi_{N_{\rm rep}}^{(\ell)})]=\pi-{\rm arg}[G''({\bm p},\tilde{\phi}_{N_{\rm rep}}^{(\ell)})]$.

The aforementioned properties of the saddle points along with the discussion of functions $G({\bm p},\phi_s)$ and $G''({\bm p},\phi_s)$
allow us to rewrite Eq.~\eqref{a29} such that
\begin{eqnarray}
{\cal A}_K^{({\rm saddle})}&=&-\ii\sqrt{\frac{\lambda}{\pi}}\Bigl(\frac{2\pi\lambda}{m\omega}\Bigr)^2\,\ee^{\ii[\Phi_0({\bm p})+\pi(N_{\rm rep}-1)F({\bm p})+\frac{1}{2}G_0({\bm p})+\frac{1}{2}\tilde{G}_0({\bm p})]}\nonumber\\
&\times&\frac{\ee^{-W({\bm p})}}{G_0''({\bm p})}\sin\Bigl[\frac{1}{2}G_0({\bm p})+\frac{1}{2}\tilde{G}_0({\bm p})-\psi_{G''}({\bm p})\Bigr]\frac{\sin[\pi N_{\rm rep}F(\bm p)]}{\sin[\pi F({\bm p})]}.
\label{a36}
\end{eqnarray}
This formula is factorized into three essential parts. It contains a term $\ee^{-W({\bm p})}/G''({\bm p})$ which is responsible for an exponential decay of the probability
amplitude of ionization ${\cal A}_K^{({\rm saddle})}({\bm p})$ while increasing the photoelectron energy. Another factor, 
$\sin\bigl[\frac{1}{2}G_0({\bm p})+\frac{1}{2}\tilde{G}_0({\bm p})-\psi_{G''}({\bm p})\bigr]$, corresponds to slow modulations of the probability
amplitude on the electron energy scale. This is in contrast to the last term in Eq.~\eqref{a36}, $\sin[\pi N_{\rm rep}F(\bm p)]/\sin[\pi F({\bm p})]$, 
which we call the {\it diffraction term}. As we are going to illustrate, this term is a source of very sharp peaks in the energy spectrum of photoelectrons,
similar to diffraction fringes observed in the experiment by Davisson and Germer~\cite{Davisson}.

\begin{figure}
\includegraphics[width=7cm]{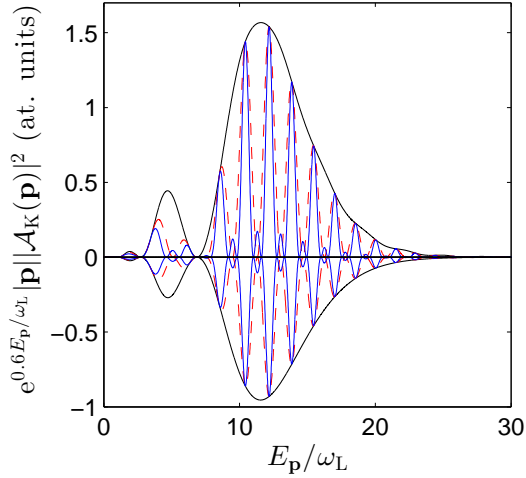}%
\caption{(Color online) Shows the energy spectra of photoelectrons~\eqref{a37} ionized by the pulse with the sin$^2$ envelope~\eqref{a12}.
The frequency of the laser field is taken $\omega_{\rm L}=1.55$ eV and its mean intensity is $I=3.125\times 10^{13}$ W/cm$^2$. The envelope of the spectra
(solid black line) corresponds to a one-cycle driving pulse ($N_{\rm rep}=1$). Other curves correspond to a sequence of either two one-cycle ($N_{\rm rep}=2$; dashed red line)
or three one-cycle ($N_{\rm rep}=3$; solid blue line) driving pulses. All results have been divided by $N_{\rm rep}^2$. They have also been multiplied by $\ee^{0.6E_{\bm p}/\omega_{\rm L}}$
to magnify the main features of the distributions. The results in the upper frame have been obtained by performing 
the integral in Eq.~\eqref{a22} exactly. The results in the lower frame have been obtained using the saddle point method~\eqref{a29}.
The major features of the mirror-reflected distributions are the same, except that they differ in magnitude. 
\label{skieldyshintandsingular20150106}}
\end{figure}

\subsubsection{Numerical illustrations}
\label{illustrations}
\begin{figure}
\includegraphics[width=7.5cm]{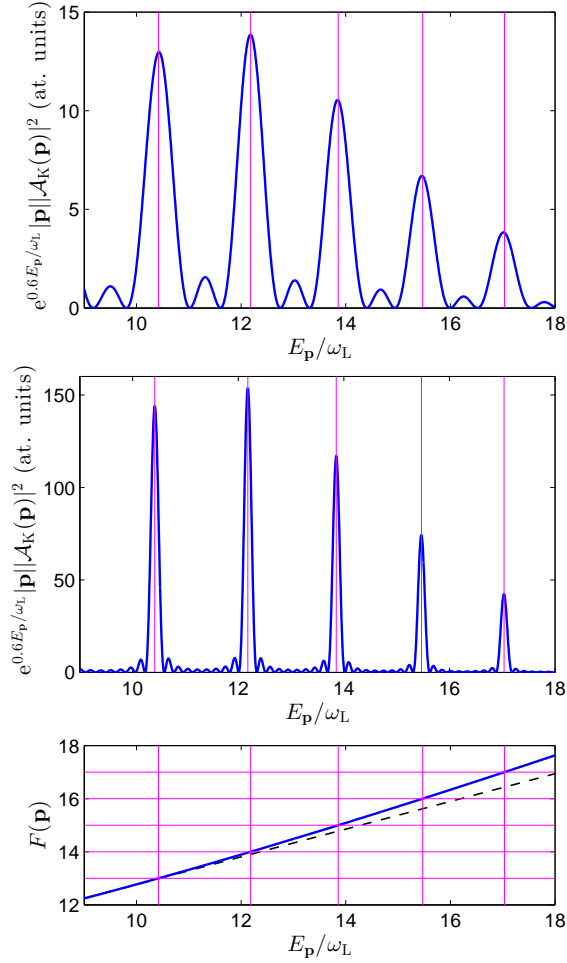}%
\caption{(Color online) In the top panel, we show a portion of the energy spectrum presented in the upper frame of Fig.~\ref{skieldyshintandsingular20150106}. Only the curve for $N_{\rm rep}=3$
is plotted, and the results are not scaled by $N_{\rm rep}^2$. Vertical lines mark the energies at which we observe the main maxima. The same but for $N_{\rm rep}=10$ 
is plotted in the middle panel. Note that in both cases the main maxima occur at the exact same energies of the final electron. At those energies the function $F({\bm p})$, drawn in the bottom panel
as the solid blue line, takes on integer values. Note that $F({\bm p})$, in contrast to the dashed black line, is not a linear function of its argument. Therefore,
the peaks in the energy distribution of photoelectrons are not equally spaced.
\label{skeldyshphasecomb20150106}}
\end{figure}

In Fig.~\ref{skieldyshintandsingular20150106} we plot the quantity $|{\bm p}||{\cal A}_K({\bm p})|^2$ which, according to Eq.~\eqref{a23}, is proportional to
a triply differential probability distribution of ionization,
\begin{equation}
|{\bm p}||{\cal A}_K({\bm p})|^2\sim \frac{{\rm d}^3{\cal P}_K}{{\rm d}\Omega_{\bm p}{\rm d}E_{\bm p}}.
\label{a37}
\end{equation}
For a visual purpose, we have multiplied this distribution by $\ee^{0.6E_{\bm p}/\omega_{\rm L}}$. While this distribution is invariant with respect to the azimuthal angle,
for the polar angle we have chosen $\theta_{\bm p}=0.2\pi$. In the upper frame, we present the exact results based on a direct numerical calculation of the integral in Eq.~\eqref{a22}.
The mirror-reflected curves, shown in the lower frame, have been calculated using the saddle point method with respect to the aforementioned integral, i.e., based on Eq.~\eqref{a29}.
While spectra in both frames differ in magnitude, their actual patterns are the same. In each frame we present three curves. The solid black envelopes correspond to the case when 
the driving pulse is a single-cycle pulse ($N_{\rm rep}=1$). As it follows from the saddle point treatment~\eqref{a36}, in this case the diffraction term equals 1 and, 
therefore, only slow modulations of the spectra are manifested. As we have also checked, for more energetic photoelectrons we observe similar modulations which, however, decrease in magnitude.
Such a behavior can be explained by the exponentially decaying term in Eq.~\eqref{a36}. The dashed red line is for $N_{\rm rep}=2$, meaning that the driving pulse
consists of two one-cycle pulses. Already in this case, a diffraction pattern is observed. We see a very sharp peaks within the envelope. This happens for any $N_{\rm rep}\geqslant 2$,
in agreement with formula~\eqref{a36}. For instance, for $N_{\rm rep}=3$, the corresponding sharp peaks are plotted with the solid blue line. Note that each spectrum was
divided by $N_{\rm rep}^2$. This resulted in nearly same heights of the peaks for different $N_{\rm rep}$. While for energies $6\omega_{\rm L}\lesssim E_{\bm p}\lesssim 25\omega_{\rm L}$, 
one can actually see that the scaled peaks have the same heights for different $N_{\rm rep}$, for energies $3\omega_{\rm L}\lesssim E_{\bm p}\lesssim 6\omega_{\rm L}$ this is not exactly the case. 
Such a behavior of the presented spectra
can be explained using the derivation based on the saddle point approximation~\eqref{a36}. According to this formula, the sharp peaks appear at  
electron energies such that $F({\bm p})=L$, where $L$ is integer. This behavior is distorted by the term $\sin\bigl[\frac{1}{2}G_0({\bm p})+\frac{1}{2}\tilde{G}_0({\bm p})-\psi_{G''}({\bm p})\bigr]$,
which manifests strongly for $3\omega_{\rm L}\lesssim E_{\bm p}\lesssim 6\omega_{\rm L}$. Let us also note that the spectra divided by $N_{\rm rep}^2$
have contact points at such electron energies $E_{\bm p}$ that the phase of the probability amplitude of ionization takes the same values regardless of $N_{\rm rep}$.

In the top panel of Fig.~\ref{skeldyshphasecomb20150106}, we plot a portion of the spectrum presented in Fig.~\ref{skieldyshintandsingular20150106} for $N_{\rm rep}=3$. The same but for
$N_{\rm rep}=10$ is plotted in the middle panel. Note that in both cases we observe the enhancement of the spectra at the exact same electron energies,
as indicated by the solid vertical lines. In the bottom panel, we show the function $F({\bm p})$ (solid blue line). As expected, the main maxima in the upper panels occur at 
those photoelectron energies when $F({\bm p})$ takes integer values. At these energies, the diffraction term in~\eqref{a36} tends to $N_{\rm rep}$ and, hence, 
the respective probability distributions scale as $N_{\rm rep}^2$. Since the major peaks become more narrow with increasing $N_{\rm rep}$,
the angle-resolved probability of ionization, when integrated over the electron energy, scales approximately as $N_{\rm rep}$. Therefore, for the Keldysh theory it is meaningful to talk about
the probability rate of ionization per one modulation of the laser pulse. Also, note that $F({\bm p})$ is not a linear function of the photoelectron kinetic
energy, which is in contrast to a straight line (dashed black line) shown in the bottom panel as well. It means that, in general, the enhancement peaks 
are not equally spaced on the photoelectron energy scale. 
Another feature which can be observed in Figs.~\ref{skieldyshintandsingular20150106} 
and~\ref{skeldyshphasecomb20150106} is that with increasing $N_{\rm rep}$, there appear $(N_{\rm rep}-2)$ additional maxima 
between any two consecutive main peaks. Their positions can be derived from Eq.~\eqref{a36}, $F({\bm p})=L+(M+1/2)/N_{\rm rep}$ where $M=1,2,...,N_{\rm rep}-2$.
These additional maxima are accompanied by zeros in the energy spectra. For $N_{\rm rep}\geqslant 2$, there is always $(N_{\rm rep}-1)$ zeros which are observed when
$F({\bm p})=L+M/N_{\rm rep}$ with $M=1,2,...,N_{\rm rep}-1$.

The diffraction pattern in the photoelectron energy spectra is observed only when $N_{\rm rep}\geqslant 2$, i.e., when the driving pulse is composed of at least two modulations.
Its features can be explained based on an approximate formula for the probability amplitude of ionization~\eqref{a36}, which suggests a very intuitive interpretation of the observed pattern.
Namely, the probability amplitudes from each modulation interfere constructively, leading to enhancements at certain electron energies. One can conclude, therefore, that each modulation acts as a slit 
in the Young-type experiment of matter waves performed by Davisson and Germer~\cite{Davisson}. Note that similar diffraction patterns can be observed in other strong-field processes as well, 
with the most recent examples in the area of strong-field quantum and classical electrodynamics~\cite{KK1,KK2,KK3,KK4} or in optics for electromagnetic waves passing through diffraction gratings~\cite{Born}.

\subsection{Combs in the GEA}
\label{combs-gea}

We have shown in the previous Section that diffraction patterns in the photoelectron energy spectra follow from the Keldysh theory. Since the Keldysh theory neglects the 
Coulomb interaction between the ejected electron and the residual ion, the question arises whether the similar patterns can be still observed if the Coulomb interaction
between the two is taken into account. To answer this question we will use now the GEA.

The probability amplitude of ionization in the first order of eikonal perturbation theory~\eqref{t51}, contains the extra space- and time-dependent phase factor
as compared to the Keldysh amplitude~\eqref{a17}. This factor functionally depends on the real-time classical trajectory in the laser field, i.e.,
\begin{equation}
{\cal A}^{(1)}({\bm p})=-\ii\frac{\ee^{\ii\Phi_0({\bm p})}}{\omega}\int_0^{2\pi}{\rm d}\phi\,\ee^{\ii G({\bm p},\phi)}\int{\rm d}^3r'\ee^{-\ii{\bm q}(\phi)\cdot{\bm r}'}(-e\bm{\mathcal{E}}(\phi)\cdot\bm{r}')\psi_0(\bm{r}')\ee^{-\ii U[\bm{r}',\phi,\bm{p}|\bm{r}_{\mathrm{cl}}]},
\label{cc1}
\end{equation}
where
\begin{equation}
U[\bm{r}',\phi,\bm{p}|\bm{r}_{\mathrm{cl}}]=-\frac{Z\alpha c}{\omega}\int_{\phi}^{2\pi}\mathrm{d}\sigma \frac{1}{|\bm{r}_{\mathrm{cl}}(\sigma;\bm{r}',\phi,\bm{p})|}\mathrm{erf}\Bigl(\sqrt{\frac{m\omega}{2\ii (\sigma-\phi)}}|\bm{r}_{\mathrm{cl}}(\sigma;\bm{r}',\phi,\bm{p})|\Bigr),
\label{cc2}
\end{equation}
and [cf. Eq.~\eqref{t45}]
\begin{equation}
\bm{r}_{\mathrm{cl}}(\sigma;\bm{r}',\phi,\bm{p})={\bm R}_{\bm p}\Bigl({\bm r}',\frac{\phi}{\omega},\frac{\sigma}{\omega}\Bigr).
\label{cc3}
\end{equation}
Note that the real-time classical trajectory $\bm{r}_{\mathrm{cl}}(\sigma;\bm{r}',\phi,\bm{p})$ depends on $\bm{r}'$ and $\bm{p}$ 
(here $\sigma$ plays the role of time in units of $1/\omega$) through the initial and final conditions, 
\begin{align}
\bm{r}_{\mathrm{cl}}(\sigma;\bm{r}',\phi,\bm{p})\Big|_{\sigma=\phi}&=\bm{r}' ,
\label{cc4a} \\
\frac{\partial}{\partial\sigma}\bm{r}_{\mathrm{cl}}(\sigma;\bm{r}',\phi,\bm{p})\Big|_{\sigma=2\pi}&=\frac{\bm{p}}{m\omega},
\label{cc4b}
\end{align}
respectively. It is worth noting that, by following the standard procedure (see, e.g., Ref.~\cite{Popruzhenko}), the functional 
\begin{equation}
W[\phi,\bm{p}|\bm{r}_{\mathrm{cl}}]=-G(\bm{p},2\pi)+G(\bm{p},\phi)-U[\bm{r}',\phi,\bm{p}|\bm{r}_{\mathrm{cl}}]
\label{cc5}
\end{equation}
can be rewritten in the form
\begin{equation}
W[\phi,\bm{p}|\bm{r}_{\mathrm{cl}}]=S[\phi|\bm{r}_{\mathrm{cl}}]+m\omega \bm{r}_{\mathrm{cl}}(\phi)\cdot \bm{r}^{\prime}_{\mathrm{cl}}(\phi)-\bm{p}\cdot \bm{r}_{\mathrm{cl}}(2\pi),
\label{cc6}
\end{equation}
where we have used the abbreviation $\bm{r}_{\mathrm{cl}}(\sigma)=\bm{r}_{\mathrm{cl}}(\sigma;\bm{r}',\phi,\bm{p})$. Here, the `\textit{prime}' means the derivative over $\sigma$ and
\begin{equation}
S[\phi|\bm{r}_{\mathrm{cl}}]=\frac{1}{\omega}\int_{\phi}^{2\pi}\mathrm{d}\sigma \mathcal{L}_{\mathrm{eff}}(\bm{r}_{\mathrm{cl}}(\sigma),\bm{r}^{\prime}_{\mathrm{cl}}(\sigma),\sigma),
\label{cc7}
\end{equation}
is the classical action with the effective Lagrangian
\begin{equation}
\mathcal{L}_{\mathrm{eff}}(\bm{r}_{\mathrm{cl}}(\sigma),\bm{r}^{\prime}_{\mathrm{cl}}(\sigma),\sigma)=\frac{m\omega^2}{2}[\bm{r}^{\prime}_{\mathrm{cl}}(\sigma)]^2+e\mathcal{E}(\sigma)\cdot \bm{r}_{\mathrm{cl}}(\sigma)-V^{(1)}_{\mathrm{eff}}(\bm{r}_{\mathrm{cl}}(\sigma),\sigma-\phi)+E_0.
\label{cc8}
\end{equation}
Here, the effective potential is defined by Eq.~\eqref{t42} with time in units of $1/\omega$. Since we deal with the static Coulomb
potential then $V_{\rm eff}^{(1)}$ depends only on the classical trajectory and the phase difference $\sigma-\phi$. Finally, the probability amplitude \eqref{cc1} can be put in the form
\begin{equation}
{\cal A}^{(1)}({\bm p})=-\ii\frac{\ee^{\ii({\bm p}^2/2m-E_0)T}}{\omega}\int_0^{2\pi}{\rm d}\phi\ \int{\rm d}^3r'\ee^{-\ii{\bm q}(\phi)\cdot{\bm r}'}(-e\bm{\mathcal{E}}(\phi)\cdot\bm{r}')\psi_0(\bm{r}')\ee^{\ii W[\phi,\bm{p}|\bm{r}_{\mathrm{cl}}]},
\label{cc9}
\end{equation}
which is suitable for the saddle point and quantum trajectory analysis. 

\subsubsection{Quantum trajectories}
\label{trajectories}

The direct integration over the space variables, ${\rm d}^3r'$, in \eqref{cc9} is very difficult to carry out as the effective potential, 
being the function of a complex argument, oscillates rapidly. For this reason, it is very convenient to apply 
the saddle point method. In our further analysis we use the simplest approximation, namely, in the effective potential we replace
the classical real-time trajectory $\bm{r}_{\mathrm{cl}}(\sigma)$ by its quantum analog (which is frequently called the complex-time trajectory~\cite{Popruzhenko})
being the solution of the free particle Newton equation in a laser field. This is in
agreement with the assumption, which is commonly made in the strong-field approximation, that the binding potential rather marginally modifies the electron trajectory in the laser field.

The quantum trajectory is the solution of the classical Newton equation in the laser field,
\begin{equation}
{\bm r}^{\prime\prime}_q(\sigma;{\bm p},\phi_s)=\frac{e}{m\omega^2}\bm{\mathcal{E}}(\sigma),
\label{q1}
\end{equation}
where the `\textit{prime}' again means the derivative with respect to the phase $\sigma$. These trajectories, however, have to fulfill the complex initial conditions for $\sigma=\phi_s$,
\begin{equation}
{\rm Re}[{\bm r}_q(\phi_s;{\bm p},\phi_s)]={\bm 0},\qquad [m\omega{\bm r}^{\prime}_q(\phi_s;{\bm p},\phi_s)]^2=-\lambda^2,
\label{q2}
\end{equation}
where $\lambda$ relates to the binding energy of a hydrogen-like atom~\eqref{a18}. Following Ref.~\cite{Popruzhenko}, we can write down 
the quantum trajectories of the form
\begin{equation}
{\bm r}_q(\sigma;{\bm p},\phi_s)=\frac{{\bm p}}{m\omega}[\sigma-{\rm Re}(\phi_s)]+\frac{e}{m}{\bm \alpha}(\sigma)-\frac{e}{m}{\rm Re}[{\bm \alpha}(\phi_s)],
\label{q3}
\end{equation}
where [cf., Eq. \eqref{a10p}]
\begin{equation}
{\bm\alpha}(\sigma)=-\frac{1}{\omega}\int_0^\sigma{\bm A}(\varphi){\rm d}\varphi=\frac{\mathcal{E}_0}{\omega^2}f_{\alpha}(\sigma)\bm{e}_z.
\label{a11}
\end{equation}
These trajectories satisfy both conditions~\eqref{q2}. The first condition in~\eqref{q2} states that the real part of the complex trajectory starts at the
center of the atom. The second condition defines the initial phase $\phi_s$, which turns out to be the saddle point solution of Eq.~\eqref{a25}.
In addition, the trajectories~\eqref{q3} are real in real phase $\sigma$,
\begin{equation}
{\rm Im}[{\bm r}_q({\rm Re}\,\sigma;{\bm p},\phi_s)]={\bm 0}.
\label{q4}
\end{equation}
The same holds for the velocity,
\begin{equation}
{\rm Im}[{\bm r}^{\prime}_q({\rm Re}\,\sigma;{\bm p},\phi_s)]={\bm 0}.
\label{q5}
\end{equation}
Since we deal with the Coulomb-free trajectories, if the laser pulse is switched off the photoelectron will carry the momentum ${\bm p}$. In other words,
\begin{equation}
m\omega{\bm r}^{\prime}_q(2\pi;{\bm p},\phi_s)={\bm p}.
\label{q6}
\end{equation}
This condition agrees with the assumption made above that the electron final state is approximated by the plane wave. It also agrees 
with the numerical analysis showing that the quantum trajectories at the end of sufficiently intense laser pulses 
(which is the case for the intensity considered in this paper) are far away from the Coulomb center. In other words, that
the electron final momentum is rather marginally affected by the interaction with the residual ion.

\begin{figure}
\includegraphics[width=7.5cm]{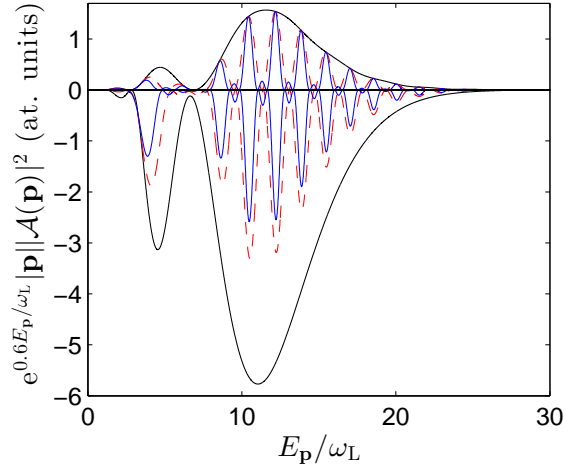}%
\caption{(Color online) The same as in Fig.~\ref{skieldyshintandsingular20150106}, except that in the lower frame the spectra have been calculated 
based on the GEA with the application of the quantum trajectory method [Eq.~\eqref{q9}].
\label{skieldysheikonal0andint}}
\end{figure}

For each saddle point $\phi_s$ and the corresponding quantum trajectory, we can define the generalized eikonal
\begin{equation}
\chi^{(1)}_q(\bm{p},\phi_s)=\frac{Z\alpha c}{\omega}\int_{\phi_s}^{2\pi}\mathrm{d}\sigma \frac{1}{|\bm{r}_q(\sigma;\bm{p},\phi_s)|}\mathrm{erf}\Bigl(\sqrt{\frac{m\omega}{2\ii (\sigma-\phi_s)}}|\bm{r}_q(\sigma;\bm{p},\phi_s)|\Bigr),
\label{q7}
\end{equation}
and its original counterpart
\begin{equation}
\chi_{q,\mathrm{original}}(\bm{p},\phi_s)=\frac{Z\alpha c}{\omega}\int_{\phi_s}^{2\pi}\mathrm{d}\sigma \frac{1}{|\bm{r}_q(\sigma;\bm{p},\phi_s)|}.
\label{q8}
\end{equation}
With these definitions the probability amplitude in the first order of eikonal perturbation theory and in the saddle-point approximation adopts the form [cf., Eq.~\eqref{a29}]
\begin{equation}
{\cal A}^{(1)}_{\rm saddle}({\bm p})=-2\sqrt{\frac{\lambda}{\pi}}\Bigl(\frac{\pi\lambda}{m\omega}\Bigr)^2\,\ee^{\ii\Phi_0({\bm p})}\sum_s\frac{\ee^{\ii G({\bm p},\phi_s)+\ii \chi^{(1)}_q(\bm{p},\phi_s)}}{G''({\bm p},\phi_s)} ,
\label{q9}
\end{equation}
and similarly for the EA, with the replacement of $\chi^{(1)}_q$ by $\chi_{q,\mathrm{original}}$.

In Fig.~\ref{skieldysheikonal0andint}, we present the energy distributions of ionized electrons similar to Fig.~\ref{skieldyshintandsingular20150106}. 
The difference is that, this time, the lower frame shows the spectra calculated within the GEA and saddle point approximation~\eqref{q9}.
In this case, we account for the Coulomb interaction between ejected photoelectrons and their parent ions. As we see in the lower frame, the positions of peaks and zeros 
in the spectra are almost identical as in the upper frame where we plot the spectra calculated based on the Keldysh approach~\eqref{a22}. What is changed, however, 
when we account for the Coulomb interaction between the electrons and the residual ions, is a significant enhancement of the ionization signal.
Also, we observe a partial loss of coherence since the distributions do not scale any longer like $N_{\mathrm{rep}}^2$.

\begin{figure}
\includegraphics[width=7cm]{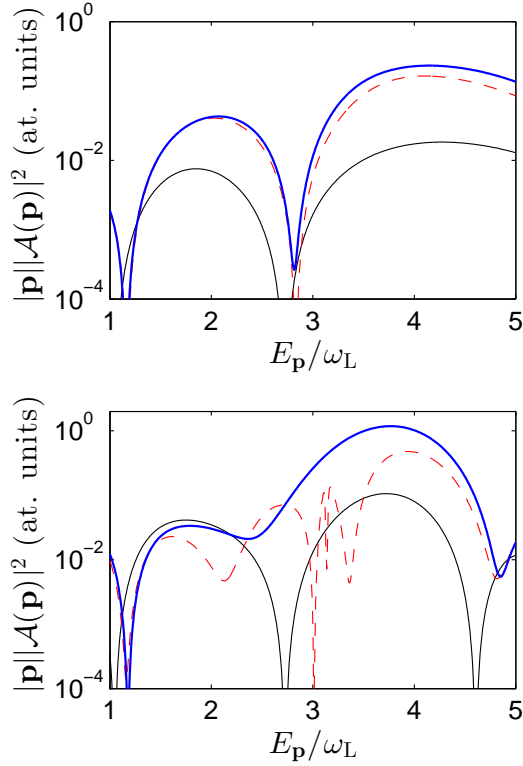}%
\caption{(Color online) 
The thin black line represents the energy distribution of ionization in the saddle point Keldysh approximation, Eq.~\eqref{a29}, 
and the thick blue line corresponds to the GEA, Eq.~\eqref{q9}. The dashed red line 
depicts the result for the EA. The upper and the lower panels show the distributions for $N_{\mathrm{rep}}=1$ or 3, respectively.
The remaining laser pulse parameters are the same as in Fig.~\ref{skieldyshintandsingular20150106}. 
\label{skieldysh1bisd20150106}}
\end{figure}

The enhancement of the ionization yield can create some doubts about the validity of eikonal perturbation theory. Let us note, however, that 
the perturbation is carried out in the exponent. In this particular case, the applicability condition for this approximation is such that in Eq.~\eqref{q9} 
the eikonal term $\chi^{(1)}(\bm{p},\phi_s)$ should be much smaller than the zeroth-order term, $G({\bm p},\phi_s)$, for both the real and imaginary parts. 
This condition is very well fulfilled for the laser pulse intensity considered in this paper.

\subsubsection{Prospects for using the GEA}
\label{prospects}

Now the question arises: To what extent the GEA is better than the EA? To answer this question, in Fig.~\ref{skieldysh1bisd20150106} we compare 
the predictions of both approaches. The thin black line represents the results calculated based on the Keldysh theory while the blue thick line is for the 
GEA. The EA results are represented by the red dashed line. In each case, the saddle point method was used.
The energy spectra presented in the upper panel are for the one-cycle pulse, $N_{\mathrm{rep}}= 1$. We observe a rather marginal difference between the GEA and EA results, 
and a significant enhancement (by roughly one order of magnitude) of these distributions as compared to the Keldysh approximation. Qualitatively,
however, all three distributions look similar. The differences appear for longer pulses, when $N_{\mathrm{rep}}> 1$. This is illustrated in the lower panel for $N_{\mathrm{rep}}=3$.
For electron kinetic energy $E_{\bm{p}}\approx 3\omega_{\mathrm{L}}$ the distributions for generalized and original eikonals differ significantly.
However, for $E_{\bm{p}}> 5\omega_{\mathrm{L}}$ (which is not shown in the figure) both approaches again give nearly the same results. The wiggles observed for the EA distribution
can be explained if we note that for the laser field parameters considered in this figure the ponderomotive energy is close to $\omega_{\mathrm{L}}$, 
which means that the structure appears for $E_{\bm{p}}\approx 3U_p$. It is well-known that for such energetic photoelectrons some of the complex trajectories can return very close to the origin 
of the Coulomb potential. As we have checked, this is the case here. Since the original eikonal is singular for such trajectories, we observe the rapid change 
of $\chi_{q,\mathrm{original}}(\bm{p},\phi_s)$ when the kinetic energy passes through the value $3U_p$ (in the considered case the real parts of $\chi_{q,\mathrm{original}}(\bm{p},\phi_s)$ 
exhibit the sharp peaks for these particular trajectories). This results in wiggles observed in the lower panel of Fig.~\ref{skieldysh1bisd20150106} for the EA. Such a behavior, however, 
is not observed for the GEA, as it is not singular for trajectories returning to the potential origin. If we compare the GEA
with the Keldysh approach, we see the enhancement of ionization but again the distributions are qualitatively similar. In our opinion, the lack of spurious behavior 
for trajectories returning back to the vicinity of the parent ion and the fact that the first Born approximation is the limiting case of the GEA
make the approach presented in this paper an attractive tool for investigations of ionization, rescattering, and high-order harmonic 
generation by strong laser pulses. This includes also more complex systems such as two-atom molecules or fullerenes.

In the literature (see, e.g., Ref.~\cite{Popruzh}) two names for the method, {\it quantum trajectories} and {\it complex-time trajectories}, are frequently used. 
In light of our analysis and more thorough studies carried out, for instance, in Refs.~\cite{Popruzh,Popruzhenko}, the second name seems to be more appropriate, 
as the trajectory ${\bm r}_q(\sigma;{\bm p},\phi_s)$ satisfies the Newton equation with the classical binding potential $V(\bm{r})$ but with the complex initial conditions. 
In such a formulation of the method, there are no quantum signatures in the definition of ${\bm r}_q(\sigma;{\bm p},\phi_s)$. This approach, however, leads 
to some problems related to the Coulomb singularity at the origin. In quantum theory (for the Schr\"odinger and Dirac equations), this singularity does not 
create any difficulties. Our investigations show that this obstacle can also be eliminated in the complex-time method. Indeed, the form 
of the effective Lagrangian~\eqref{cc8} suggests to assume that, up to the first order of eikonal perturbation theory, the trajectories should fulfill the Newton equation of the form [cf., Eq.~\eqref{q1}],
\begin{equation}
m\omega^2{\bm r}^{\prime\prime}_q(\sigma;{\bm p},\phi_s)=e\bm{\mathcal{E}}(\sigma)-\bm{\nabla}V^{(1)}_{\mathrm{eff}}({\bm r}_q(\sigma;{\bm p},\phi_s),\sigma-\phi_s),
\label{q10}
\end{equation}
with a suitable initial conditions. The effective potential $V^{(1)}_{\mathrm{eff}}(\bm{r},\sigma)$, contrary to the classical one $V(\bm{r})$, is not singular 
at the origin for non-zero time and is smeared out by the `\textit{quantum diffusion}' represented by the Laplacian and the non-linear term in 
Eqs.~\eqref{t25} or \eqref{t27}. In other words, it accounts for the spreading of the electron wave packet during the quantum time evolution.
The Laplacian introduces the Planck constant into the definition of the effective potential in the first order 
eikonal perturbation theory. In other words, $V^{(1)}_{\mathrm{eff}}(\bm{r},\sigma)$ differs from $V(\bm{r})$ by quantum corrections which vanish in the 
limit $\hbar\rightarrow 0$. This also means that ${\bm r}_q(\sigma;{\bm p},\phi_s)$ does contain quantum corrections and, therefore, we should rather call 
those trajectories `\textit{complex-time quantum trajectories}'. The effects related to the quantum corrections in ${\bm r}_q(\sigma;{\bm p},\phi_s)$ are now under investigations.

\section{Conclusions}
\label{conclusions}

We have formulated the GEA for ionization processes driven by strong laser pulses.
As we have shown, the Born approximation arises as the limiting case of our approach. The EA does not have this
property, which significantly diminishes its applicability to the rescattering phenomena. Moreover, the EA is singular for 
the trajectories that come back to the center of atomic potential. We have demonstrated that the GEA does not have this shortcoming either. This makes
it a very promising tool to study rescattering-related phenomena, with the most prominent example of high-order harmonic generation.

Using the GEA we have discussed the appearance of coherent diffraction patterns in photoelectron
energy spectra and their modifications induced by the interaction of photoelectrons with the parent ion. We have identified the 
conditions necessary to obtain such coherent patterns. If a pulse 
consists of at least two modulations, each of these modulations acts as a slit in the Young-type experiment for matter waves
resulting in a coherent enhancement of ionization signal at particular electron energies. As we have illustrated this numerically,
if we increase the number of modulations within a pulse, the comb-like structures in the energy spectrum of photoelectrons
become similar to the $\delta$-like structures. This is particularly interesting in the context of designing new sources of
electron pulses, which is another topic to be studied in near future.

\section*{Acknowledgements}
This work is supported by the Polish National Science Center (NCN) under Grant No. 2012/05/B/ST2/02547. F.C.V. acknowledges the support from the Foundation for Polish Science
International PhD Projects Programme co-financed by the EU European Regional Development Fund. Moreover, K.K. acknowledges the support from the Kosciuszko Foundation and the hospitality of the Department of
Physics and Astronomy at the University of Nebraska, Lincoln, Nebraska, where part of this paper was prepared.


\begin{thebibliography}{99}

\bibitem{Bruns}
The term ``{\it eikonal}`` was introduced by H. Bruns, in {\it Das Eikonal}, Abh. Kgl. s\"achs Ges. Wies., math-phs. Kl. {\bf 21}, 370 (1895).
From Greek it means ''{\it image}``.  

\bibitem{Born}
M. Born and M. Wolf, {\it Principles of Optics: Electromagnetic Theory of Propagation, Interference and Diffraction of Light} (Pergamon, London, 1959).

\bibitem{eikhist1}
G. Moli\`ere, Z. Naturforsch. {\bf 2}, 133 (1947).

\bibitem{Joachain}
Ch. J. Joachain, {\it Quantum Collision Theory} (North-Holland, Amsterdam, 1975).

\bibitem{bib:burke} 
P. G. Burke, \textit{Potential Scattering in Atomic Physics} (Plenum, New York, 1977).

\bibitem{Landau}
L. D. Landau and E. M. Lifschitz, {\it Quantum Mechanics: Non-Relativistic Theory} (Pergamon, Oxford, 1991).

\bibitem{eikhist2}
R. J. Glauber, in {\it Lectures in Theoretical Physics}, vol. {\bf 1}, p. 315, (ed. W. E. Brittin and L. G. Dunham) (Interscience, New York, 1959).

\bibitem{eiklas1}
B. J. Choudhury and B. S. Bakar, J. Phys. B {\bf 7}, L137 (1974); {\it ibid.} {\bf 8}, L228 (1975).

\bibitem{eiklas2}
B. A. Zon, J. Phys. B {\bf 8}, L86 (1975).

\bibitem{bib:krstic} 
P. Krsti\'{c} and M. H. Mittleman,  Phys. Rev. A {\bf 25}, 1568 (1982).

\bibitem{Jurek}
J. Z. Kami\'nski, Acta Phys. Pol. A {\bf 66}, 517 (1984). 

\bibitem{Fock}
V. A. Fock, Phys. Z. Sowjetunion {\bf 12}, 404 (1937).

\bibitem{Schwinger}
J. Schwinger, Phys. Rev. {\bf 82}, 664 (1951).

\bibitem{Schwinger2}
J. Schwinger, in {\it Selected papers on Quantum Electrodynamics}, edited by J. Schwinger, p. 664 (Dover, New York, 1958).

\bibitem{Fradkin1}
E. S. Fradkin, Nucl. Phys. {\bf 76}, 588 (1966).

\bibitem{Fradkin2}
I. A. Batalin and E. S. Fradkin, Teor. Mat. Fiz. {\bf 5}, 190 (1970).

\bibitem{Fradkin3}
E. S. Fradkin, U. Esposito, and S. Termini, Rev. Nuovo Cim. {\bf 11}, 498 (1970).

\bibitem{bib:kleber} 
M. Kleber, Phys. Rep. {\bf 236}, 331 (1994).

\bibitem{geiksch}
H. K. Avetissian, A. G. Markossian, G. F. Mkrtchian, and S. V. Movsissian, Phys. Rev. A {\bf 56}, 4905 (1997).

\bibitem{geikdirac}
H. K. Avetissian, K. Z. Hatsagortsian, A. G. Markossian, and S. V. Movsissian, Phys. Rev. A{\bf 59}, 549 (1999).

\bibitem{Reis}
H. R. Reiss and V. P. Krainov, Proc. SPIE {\bf 39}, 2796 (1996).

\bibitem{Krainov}
V. P. Krainov, J. Opt. Soc. Am. B {\bf 14}, 425 (1997).

\bibitem{Gord}
S. Gordienko and J. M. ter Vehn, Proc. SPIE {\bf 5228}, 416 (2003).

\bibitem{Goresla}
S. P. Goreslavski, G. G. Paulus, S. V. Popruzhenko, and N. I. Shvetsov-Shilovski, Phys. Rev. Lett. {\bf 93}, 233002 (2004).

\bibitem{Faisal1}
F. H. M. Faisal and G. Schlegel, J. Phys. B {\bf 38}, L223 (2005).

\bibitem{Potvliege}
C. C. Chiril\v{a} nd R. M. Potvliege, Phys. Rev. A {\bf 71}, 021402 (2005).

\bibitem{Faisal2}
F. H. M. Faisal and G. Schlegel, J. Mod. Opt. {\bf 53}, 207 (2006).

\bibitem{Thumm}
C.-H. Zhang and U. Thumm, Phys. Rev. A {\bf 82}, 043405 (2010).

\bibitem{Krai}
V. P. Krainov and B. Shokri, Zh. Eksp. Teor. Fiz. {\bf 107}, 1180 (1995).

\bibitem{Smirnova1}
O. Smirnova, A. S. Mouritzen, S Patchkovskii, and M. Yu Ivanov, J. Phys. B {\bf 40}, F197 (2007).

\bibitem{Smirnova}
O. Smirnova, M. Spanner, and M. Ivanov, Phys. Rev. A {\bf 77}, 033407 (2008).

\bibitem{Becker}
D. B. Milo\v{s}evi\'{c}, G. G. Paulus, D. Bauer, and W. Becker, J. Phys. B {\bf 39}, R203 (2006).

\bibitem{Popruzh}
S. V. Popruzhenko, J. Phys. B {\bf 47}, 204001 (2014).

\bibitem{Keldysh}
L. V. Keldysh, Sov. Phys. JETP {\bf 20}, 1307 (1965).

\bibitem{Faisal}
F. H. M. Faisal, J. Phys. B {\bf 6}, L89 (1973).

\bibitem{Reiss}
H. R. Reiss, Phys. Rev. A {\bf 22}, 1786 (1980).

\bibitem{bib:wbecker} 
W. Becker, L. Davidovich, and J. K. McIver, Phys. Rev. A, {\bf 49}, 1131 (1994).

\bibitem{Popruzhenko}
S. V. Popruzhenko and D. Bauer, J. Mod. Optics {\bf 55}, 2573 (2008).

\bibitem{Klaiber}
M. Klaiber, E. Yakaboylu, and K. Z. Hatsagortsyan, Phys. Rev. A {\bf 87}, 023417 (2013).

\bibitem{Tzoar}
M. Jain and N. Tzoar, Phys. Rev. A {\bf 18}, 538 (1978).

\bibitem{Cavaliere}
P. Cavaliere, G. Ferrante, and C. Leone, J. Phys. B {\bf 13}, 4495 (1980).

\bibitem{CK0}
J. Z. Kami\'nski, Phys. Scr. {\bf 34}, 770 (1986).

\bibitem{CK1}
J. Z. Kami\'nski, Phys. Rev. A {\bf 37}, 622 (1988).

\bibitem{CK2}
S. Basile, F. Trombetta, G. Ferrante, Phys. Rev. Lett. {\bf 61}, 2435 (1988).

\bibitem{CK3}
D. G. Arb\'o, J. Phys. B {\bf 47}, 204008 (2014).

\bibitem{Perelomov}
A. M. Perelomov and V. S. Popov, Zh. Exp. Theor. Fiz. {\bf 50}, 1393 (1966).

\bibitem{Popov1}
V. S. Popov, Phys. Usp. {\bf 47}, 855 (2004).

\bibitem{Popov2}
V. S. Popov, Phys. At. Nucl. {\bf 68}, 686 (2005)

\bibitem{Popov3}
V. S. Popov, V. D. Mur, and S. V. Popruzhenko, JETP Lett. {\bf 85}, 223 (2007).

\bibitem{Gribakin}
G. F. Gribakin and M. Yu. Kuchiev, Phys. Rev. A {\bf 55}, 3760 (1997).

\bibitem{Popruzhen}
S. V. Popruzhenko, G. G. Paulus, and D. Bauer, Phys. Rev. A {\bf 77}, 053409 (2008͒).

\bibitem{eikHeidelberg}
M. Klaiber, E. Yakaboylu, and K. Z. Hatsagortsyan, Phys. Rev. A {\bf 87}, 023418 (2013).

\bibitem{Fetter} 
A. L. Fetter and J. D. Walecka, {\it Quantum Theory of Many Particle Systems} (McGraw-Hill, New York, 1971).

\bibitem{Dirac}
P. A. M. Dirac, Phys. Z. Sowjetunion {\bf 3}, 64 (1933).

\bibitem{Dirac2}
P. A. M. Dirac, in {\it Selected papers on Quantum Electrodynamics}, edited by J. Schwinger, p. 312 (Dover, New York, 1958).

\bibitem{FeynmanHibbs}
R. P. Feynman and A. R. Hibbs, {\it Quantum Mechanics and Path Integrals} (McGraw-Hill, New York, 1965).

\bibitem{Davisson}
C. J. Davisson and L. H. Germer, Phys. Rev. {\bf 30}, 705 (1927).

\bibitem{KK1}
K. Krajewska and J. Z. Kami\'nski, Laser Phys. Lett. {\bf 11}, 035301 (2014).

\bibitem{KK2}
K. Krajewska, M. Twardy, and J. Z. Kami\'nski, Phys. Rev. A {\bf 89}, 052123 (2014).

\bibitem{KK3}
K. Krajewska and J. Z. Kami\'nski, Phys. Rev. A {\bf 90}, 052108 (2014).

\bibitem{KK4}
K. Krajewska and J. Z. Kami\'nski, Proc. SPIE {\bf 9441}, 94410S (2014).


\end{thebibliography}
\end{document}